\documentstyle[pra,aps]{revtex} 
\begin{document} 
  \draft 
  \title{Field theory model giving rise to  
 "quintessential inflation"  \\
without the cosmological constant
and other fine tuning problems} 
\author{A.  B.  Kaganovich \thanks{alexk@bgumail.bgu.ac.il}} 
\address{Physics Department, Ben Gurion University of the Negev, Beer Sheva 
84105, Israel}
\maketitle
\begin{abstract} 
A field theory is developed based on the idea that the effective
action of yet unknown fundamental theory, at energy scale below the
Planck mass $M_{p}$
has the form of expansion in two measures: $S=\int d^{4}x[\Phi L_{1}+
\sqrt{-g}L_{2}]$ where the new measure $\Phi$ is defined using the 
antisymmetric tensor field 
$\Phi d^4x=\partial_{[\alpha}A_{\beta\gamma\delta]}dx^\alpha\wedge
dx^\beta\wedge dx^\gamma\wedge dx^\delta$ . A shift $L_{1}\rightarrow
L_{1}+const$  does not affect the equations of motion whereas a similar shift 
when
implementing with $L_{2}$ causes  a change which in the standard GR would be
equivalent to that of the cosmological constant (CC) term.  The next basic
conjecture is that the Lagrangian densities
$L_1$ and $L_2$ do not depend of $A_{\mu\nu\lambda}$.
The new measure degrees of freedom result in the scalar field $\chi =
\Phi /\sqrt{-g}$ alone.  The  constraint appears that determines $\chi$ in 
terms of matter fields. After the conformal transformation to the new variables
(Einstein frame),
all equations of motion take the canonical GR form of the equations for
gravity and matter fields and, therefore the models we study are free of 
the well-known "defects" that distinguish the Brans-Dicke type theories from
Einstein's GR.  
All novelty is revealed only in an unusual structure of the effective
potentials and interactions which turns over our intuitive
ideas based on our experience in field theory. For example, the greater 
$\Lambda$ we
admit in $L_{2}$, the smaller magnitude of the effective inflaton potential 
$U(\phi)$ 
will be in the Einstein picture. Field
theory models are suggested with explicitly broken global continuous
 symmetry, which in the Einstein frame has the form 
$\phi\rightarrow\phi+const$. The symmetry  restoration occurs 
as $\phi\rightarrow\infty$.
A few models are presented where the effective potential $U(\phi)$ is
produced with the following shape:
for $\phi\lesssim -M_{p}$, $U(\phi)$ has the form typical for  
inflation model,
e. g. $U=\lambda\phi^4$ with $\lambda\sim 10^{-14}$; \, for 
$\phi \gtrsim -M_{p}$, $U(\phi)$ has mainly the exponential form
$U\sim e^{-a\phi/M_{p}}$ with  variable $a$:
   \, $a=14$ for $-M_{p}\gtrsim \phi \lesssim M_{p}$,  that gives
the possibility for nucleosynthesis and large-scale structure formation;
   \, $a=2$  for $\phi \gtrsim M_{p}$, that implies the quintessence era.
There is
no need of any fine tuning to prevent appearance of the CC term or any other
terms that could violate the flatness of $U(\phi)$  at 
$\phi\gg M_{p}$. \, $\lambda\sim 10^{-14}$ is obtained without fine tuning
as well. 
Quantized matter fields models, including spontaneously broken gauge 
theories, can be incorporated without altering the mentioned above results.
Direct coupling of fermions to the inflaton resembles
the Wetterich's  model but there is a possibility to avoid
any observable  effect at the late universe.
SSB does not raise any problems with the CC in the late universe.    
    
\end{abstract}          
    
    \renewcommand{\baselinestretch}{1.6} PACS number(s): 11.15.Ex, 
98.80.Cq, 
12.10.Dm, 04.90.+e 

\pagebreak 

\section{Introduction} 
Recent 
high-redshift and CMB data \cite{P} suggest that a small effective
 cosmological constant
gives a dominant contribution to the energy density 
of the present universe.  Among the attempts to describe this 
picture, the idea to profit by the properties of a slow-rolling 
scalar field (quintessence model) \cite{Wett1988NP668}
-\cite{Stein} 
seems to be the most attractive and successful. In such approach, the 
present vacuum energy density 
$\rho_{vac}\sim 10^{-47}(GeV)^4$
has to  be imitated by the energy density of a slowly rolling scalar 
field down its 
potential $U(\phi)$
 which presumably approaches zero as $\phi\rightarrow\infty$.
However all known quintessence models contain two fundamental 
problems:

1. The cosmological constant problem\cite{CCP}, \cite{Star} remains in the 
quintessence models as well: particle physics and cosmology must give a 
distinct
mechanism that enforces the effective cosmological constant to decay
from an extremely large value in the very early universe
to the extremely small present value without fine tuning of parameters
and initial conditions.

2. All known quintessence models are based on the {\em choice} of some
specific form for the potential $U(\phi)$. The general feature 
of the potentials needed to realize quintessence is that $U(\phi)$
must be flat enough as $\phi$
is large enough in order to provide conditions for 
the slow-roll approximation. However it is not clear what happens 
with other possible terms in the potential, including quantum 
corrections (see Kolda and Lyth, \cite{KLyth}). In fact, the potential 
may for instance contain
terms that constitute a
structure of polynomials in $\phi$ (and $\phi^{n}\ln \phi$) 
 and they are not negligible as $\phi$ is large, unless
an extreme fine tuning is assumed for the mass and
self-couplings.
For example, the restriction of the  flatness
conditions  on the quartic self-interaction  $\lambda \phi^4$ 
is\cite{KLyth} 
$\lambda\ll
10^{-120}(\frac{M_{p}}{\phi})^{2}$.

In this paper I am going to present a field theory model that 
resolves
the above fine tuning problems and besides that, this model is able 
to give a broad range of tools for constructive  
 answers few more important questions:

3. In the framework of a model
where potential $U(\phi)$ of the exponential or inverse power low 
(or there combinations\cite{Stein}) form plays the role of a 
quintessential potential as $\phi$ is large enough, the
 question arises what 
is the cosmological
role of such $U(\phi)$ as $\phi$ is close to zero or negative. If some 
other scalar field 
 is responsible for an inflation of the early universe,
 then a field theory has to explain why
the  potential $U(\phi)$ of the scalar field $\phi$ is 
negligible
as $\phi$ is close to zero or negative. However, if the same quintessence 
field $\phi$
 plays also the  role of
the inflaton\cite{Sp}, \cite{PV} (in  the  early universe) then again a
 field theory
 has to explain\cite{Rosati} an origin 
 of the relevant effective potential. Of
course this is a  nontrivial problem. For example,
Peebles and Vilenkin\cite{PV} have presented an interesting model 
of a single 
scalar field that drives the 
inflation of the early universe and ends up as quintessence. 
They adopt the monotonic potential
\begin{eqnarray}
U(\phi)&=&\lambda m^4\left[1+(\phi/m)^4\right]\qquad for\qquad \phi<0,
                          \nonumber\\
   &&=\frac{\lambda m^4}{1+(\phi/m)^{\alpha}}\qquad for\qquad \phi\geq 0
\label{peebvilen}
\end{eqnarray}
where $\alpha =const>0$ (for example 4 or 6) and
 the parameters $\lambda = 10^{-14}$ and $m=8\times 10^{5}GeV$
were adjusted in \cite{PV}  to achieve a satisfactory
 agreement with the main observational constraints. It is well 
known\cite{Linde1} that such  extremely small value of 
$\lambda$ is dictated in the $\lambda \phi^4$ theory of the chaotic 
inflation scenario by the necessity to obtain a density 
perturbation $\frac{\delta\rho}{\rho}\sim 10^{-5}$ in the 
observable part of the universe. In other words, the potential of this
"quintessential inflation" model includes {\em both} the fine tuning 
required by the inflation of the early universe  {\em and}  the 
fine tuning dictated by the quintessence model of the late universe. 
As it is pointed out in Ref.\cite{PV}, it seems also 
to be an unnatural feature of this model that a small mass 
$m=8\times 10^{5}GeV\ll M_{p}$ must appear in the potential of the 
inflaton field $\phi$ interacting only with gravity. And finally,
one should apparently believe that such sort of 
"quintessential inflation" potential must be generated by some 
field theory without fine tuning. These problems are typical for the
quintessential inflation type models\cite{Sp}, \cite{PV} .

4. It is well known that
 the " coincidence problem"
\cite{Stein1} can be avoided in the framework
of the quintessence models that make use "tracker potentials"\cite{Stein}.
The exponential potential with $a= const$
\begin{equation}     
U(\phi)=U_{0}e^{-a\phi/M_{p}} 
\label{expintrod}
   \end{equation}
is a special example of a  tracker solution\cite{Stein}. 
In  spacially flat  models with such potential,
the ratio of the scalar field 
$\phi$ energy density to the total matter energy density rapidly
approaches a constant value determined by $a$ and the matter equation
of state\cite{Wett1988NP668},  \cite{RatraP},  \cite{Ferr-Joyce}
(see also
Ref. \cite{Gkaluza} where a similar result was achieved in the context of 
Kaluza-Klein-Casimir cosmology). However the strong 
constraint on $\Omega_{\phi}$  dictated by cosmological nucleosynthesis
($\Omega_{\phi}\lesssim 0.2$ )\cite{Wett}, \cite{Ferr-Joyce}, 
\cite{Birkel} predetermines for the $\phi$-fraction to remain subdominant 
one in the
future that apparently contradicts the observable accelerated expansion.
A  possible
resolution of this problem  proposed by Wetterich\cite{Wett} consists in the 
idea  that $a$ in (\ref{expintrod}) might be $\phi$-dependent. 
In that case it would be again very attractive to develop a  field theory model
where the exponential potential (\ref{expintrod}) with an appropriate
$\phi$-dependent $a$ is generated in a natural way.

5. Since the mass of excitations of the $\phi$-field has to be 
extremely small in the present-day universe
 ($m_{\phi}\leq H_{0}\sim 10^{-33}eV$),
possible direct couplings 
of  $\phi$ to the 
standard matter fields should give rise to  very long-range forces 
which do 
 not  obey the equivalence principle\cite{Carroll}. To prevent such 
undesirable
effects, the very strong  upper limits on the coupling constants 
of the 
quintessence field to the 
standard matter fields have to be accepted without any known reason:
an attempt to construct a model where an unbroken symmetry could support
zero mass of $\phi$-excitations\cite{Dolgov} inevitably runs against the 
necessity to start from  a trivial 
potential\cite{Carroll}; without knowledge of a mechanism for the 
breaking of this symmetry, such small coupling constants may be 
introduced into a theory only by hand.

It will be shown in this paper that one can  answer all the above 
questions 1-5  in the framework of the field theory model 
 based on the hypothesis that the effective action of the 
fundamental theory at the energy scales  below the Planck mass
can be represented in
a general form including {\em two measures and 
respectively, two Lagrangian densities }
\begin{equation}
    S = \int\left[\Phi L_{1}+\sqrt{-g}L_{2}\right]d^{4}x 
\label{S} 
\end{equation}

Here $\sqrt{-g}$ is the standard measure of integration in the action
principle of Einstein's General Relativity (GR) and other gravitational
 theories making use the general coordinate invariance.
 The measure $\Phi$ is 
defined using the antisymmetric tensor field $A_{\mu\nu\lambda}$
\begin{equation}
\Phi d^4x=\partial_{[\alpha}A_{\beta\gamma\delta]}dx^\alpha\wedge
dx^\beta\wedge dx^\gamma\wedge dx^\delta 
\label{Fi} 
\end{equation}
and (\ref{S}) is also invariant under general coordinate 
transformations. Notice that the measure 
$\Phi$ is a total derivative and therefore a shift $L_{1}\rightarrow 
L_{1}+const$  does not affect 
equations of motion whereas a 
similar shift when implementing with $L_{2}$ 
 causes  a change which in the standard GR would be
 equivalent to that of the cosmological constant term.  
 The next basic conjecture is that the Lagrangian 
densities $L_1$ and
$L_2$ do not depend of $A_{\mu\nu\lambda}$. In this paper I refer to 
this theory as the "two measures theory" (TMT) .

The main features of 
TMT have been studied in series of papers\cite{GK1}-\cite{G1}

\section{Some general features of TMT } 

Let us consider a 
simple model with the scalar field $\phi$ 
\begin{equation}
S=\int d^{4}x\left[\Phi\left(
-\frac{1}{\kappa}R(\Gamma ,g)+\frac{1}{2}g^{\mu\nu}\phi_{,\mu} 
\phi_{,\nu}-V_{1}(\phi)\right)+\sqrt{-g}V_{2}(\phi)\right] 
\label{totac} 
\end{equation} 

The case where $V_{2}(\phi)\equiv const$ was studied in Ref. \cite{GK4}
and the general case was studied by Guendelman in Ref.
\cite{G}. 
TMT gives desirable results 
if we proceed in the first order formalism (metric $g_{\mu\nu}$ and 
connection $\Gamma^{\mu}_{\lambda\sigma}$ are independent variables as well 
as the antisymmetric tensor field $A_{\mu\nu\lambda}$)
and 
$R(\Gamma,g)=g^{\mu\nu}R_{\mu\nu}(\Gamma)$,\hspace{0.25cm} 
$R_{\mu\nu}(\Gamma)=R^{\alpha}_{\mu\nu\alpha}(\Gamma)$ and 
\begin{equation} 
R^{\lambda}_{\mu\nu\sigma}(\Gamma)\equiv \Gamma^{\lambda}_{\mu\nu,\sigma}+ 
\Gamma^{\lambda}_{\alpha\sigma}\Gamma^{\alpha}_{\mu\nu}- 
(\nu\leftrightarrow\sigma).  
\label{R4}
\end{equation}
At this stage no 
specific form for $V_{1}(\phi)$ and $V_{2}(\phi)$ is assumed. 

Variation of the action  with respect to $A_{\mu\nu\lambda}$
results in 
equation $\epsilon^{\mu\nu\alpha\beta}\partial_{\beta}L_{1}=0$
which means that
\begin{equation} 
L_{1}=-\frac{1}{\kappa}R(\Gamma,g)+\frac{1}{2}g^{\mu\nu}\phi_{,\mu} 
\phi_{,\nu}-V_{1}(\phi)=sM^{4}=const ,
\label{L1M} 
\end{equation}
where $sM^4$ is 
an integration constant, $s=\pm 1$ and $M$ is a constant of the dimension 
of mass. 

Variation with respect to 
$g^{\mu\nu}$ leads to 
\begin{equation} 
-\frac{1}{\kappa}R_{\mu\nu}({\Gamma})+\frac{1}{2}\phi_{,\mu}\phi_{,\nu} 
-\frac{1}{2\chi}V_{2}(\phi)g_{\mu\nu}=0 
\label{Rmn} 
\end{equation} 
where 
the scalar field $\chi$ is defined by
\begin{equation} 
\chi\equiv\frac{\Phi}{\sqrt{-g}} 
\label{chi} 
\end{equation}

Consistency 
condition of equations (\ref{L1M}) and (\ref{Rmn}) takes the form of 
the 
{\em constraint}
\begin{equation} 
    V_{1}(\phi)+sM^{4}-\frac{2V_{2}(\phi)}{\chi}=0 
\label{c}
\end{equation}

Solution of equations obtained by variation of the action with respect to 
$\Gamma^{\mu}_{\lambda\sigma}$ can be represented (see \cite{GK2} -
 \cite{GK4}) as a sum of 
 the Christoffel's connection 
coefficients $\{ ^{\lambda}_{\mu\nu}\}$ of the metric $g_{\mu\nu}$ and 
a non-Riemannian part which is a linear combination of $\sigma,_{\mu}$
where
$\sigma\equiv\ln\chi$.  

The scalar field $\phi$ equation is 
\begin{equation} 
(-g)^{-1/2}\partial_{\mu}(\sqrt{-g}g^{\mu\nu}\partial_{\nu}\phi)+ 
\sigma,_{\mu}\phi 
^{,\mu}+\frac{dV_{1}}{d\phi}-\frac{1}{\chi}\frac{dV_{2}}{d\phi}=0, 
\label{SE}
\end{equation}

In the conformal frame defined by the conformal 
transformation 
\begin{equation} 
{g}_{\mu\nu}(x)\rightarrow g^{\prime}_{\mu\nu}(x)=\chi g_{\mu\nu}(x); 
\qquad \phi\rightarrow \phi; \qquad A_{\mu\nu\lambda}\rightarrow
A_{\mu\nu\lambda} 
\label{ct} 
\end{equation} 
the non-Riemannian contribution 
into the connection disappears: 
${\Gamma}^{\lambda}_{\mu\nu}\rightarrow\overline{\Gamma}^{\lambda}_{\mu\nu}= 
\overline{\{ ^{\lambda}_{\mu\nu}\}}$ \qquad (here \qquad $\overline{\{ 
^{\lambda}_{\mu\nu}\}}$ are the Christoffel's connection coefficients of 
the Riemannian space-time with the metric $g^{\prime}_{\mu\nu}$).  
Tensors $R^{\lambda}_{\mu\nu\sigma}(\Gamma)$ and $R_{\mu\nu}(\Gamma)$ 
transform to the Riemann 
$R^{\lambda}_{\mu\nu\sigma}(g^{\prime}_{\alpha\beta})$ and Ricci 
$R_{\mu\nu}(g^{\prime}_{\alpha\beta})$ tensors respectively in the 
Riemannian space-time with the metric $g^{\prime}_{\mu\nu}$.  After
making use the 
solution for $\chi$ as it follows from the constraint (\ref{c}), the 
gravitational equations (\ref{Rmn}) and the scalar 
field equation (\ref{SE}) in the new conformal frame obtain {\em the standard 
form of the Einstein's GR equations for the selfconsistent system of 
gravity 
($g^{\prime}_{\mu\nu}$) and scalar field $\phi$ with the TMT effective  
potential} (for details see \cite{GK4} and \cite{G1})
\begin{equation}
    U(\phi) = \frac{1}{\chi^{2}}V_{2}(\phi)= 
\frac{1}{4V_{2}(\phi)}[sM^4 +V_{1}(\phi)]^{2} 
\label{VW}
\end{equation} 

Notice that just $U(\phi)$ plays the role of the 
true potential that governs the dynamics of the scalar field $\phi$ 
while $V_{1}(\phi)$ and $V_{2}(\phi)$ have no sense of the potential 
energy densities themselves but rather they generate the potential 
energy density. This is why we will  use the term 
{\em pre-potentials} for
$V_{1}(\phi)$ and $V_{2}(\phi)$. Notice that our choice of the sign in 
front of the pre-potential $V_{2}(\phi)$ is opposite to the usual one 
that would be in the case of the standard GR. This is doing just  
for convenience in what follows.

In order to provide disappearance of  the
cosmological constant, one {\em demands} usually that the effective 
potential is equal to zero at the  minimum, i.e. it is 
necessary that the effective potential and its  first derivative
are equal to zero at the same point. As a matter of fact this is the
essence of  the cosmological 
constant problem  treated in the "old" sense when
there was no  need in explanation of a small but
non-zero cosmological constant at present.
If we want to avoid the necessity to 
fulfill this fine tuning, TMT gives us such an opportunity and it has 
been explored in Refs. \cite{GK4}, \cite{G1}).  
In fact, {\em independently of the shape} of the 
nontrivial pre-potential $V_{1}(\phi)$, {\em infinite number of initial 
conditions} exists for which $V_{1}+sM^{4}=0$ at some value  $\phi
=\phi_{0}$. If $V_{1}(\phi)$ and $V_{2}(\phi)$ are regular at $\phi
=\phi_{0}$, $V_{1}^{\prime}(\phi_{0})\neq 0$ and
$V_{2}(\phi)$ is positive definite then  
$\phi=\phi_{0}$ is the absolute minimum of $U(\phi)$ with the value
$U(\phi_{0})=0$ . We will refer to such a situation as {\em the first class
scenario}. 

 In the present paper we will study the models with
such a pre-potential $V_{1}$ that there will be
  infinite number of initial 
conditions for which $V_1+sM^4\neq 0$ at any value of $\phi$ (we will
refer to such a situation as {\em the second class  scenario} ). Then the
stable vacuum may for instance be realized asymptotically as 
$\phi\rightarrow\infty$, which is actually the idea used in the 
quintessence models.   

The assumption that $V_{2}(\phi)$ is positive definite
will be our choice in what follows.

\section{Extremely broad  class  of TMT  models  does not require fine
tuning\\ to provide quintessence.}

\subsection{General idea: the inverse power low quintessential potential
as a simple example.}
In contrast to  standard gravitational theories where the quintessential 
potential must be
slow {\em decreasing} function as $\phi\rightarrow\infty$, in TMT
we have an {\em absolutely new option}: the quintessential behaviour of the 
TMT effective potential $U(\phi)$ for $\phi$ large enough may be achieved 
with {\em increasing} prepotentials $V_{1}(\phi)$ and $V_{2}(\phi)$. This
circumstance enables to avoid both the cosmological constant problem and
the problem of the flatness of the quintessential potential.

For illustration of these statements let us notice that starting from  
the positive power low pre-potentials $V_{1}$ and $V_{2}$ 
\begin{equation} 
V_{1}  =  m_{1}^{(4-n_{1})} \phi^{n_{1}};  \qquad
V_{2}  = \frac{1}{4} m_{2}^{(4-2n_{2})} \phi^{2n_{2}}. 
\label{V12poz}
\end{equation}
with $n_2>n_1$, we obtain the TMT effective potential which for $\phi$ 
large enough has the inverse power low form
\begin{equation}
U \approx \frac{m_{1}^{2(4-n_{1})}}{m_{2}^{2(2-n_{2})}} 
\frac{1}{\phi^{2(n_2 -n_1)}}
\label{U5}
\end{equation}
and does not depend on the integration constant. Another interesting 
 case is $V_{1}\equiv 0$ (remined that adding  a 
constant to $V_{1}$ is equivalent
 just to a redefinition of the integration constant $sM^4$) and, for example,
$V_2\equiv \lambda\phi^4$. Then $U(\phi)=M^{8}/\lambda\phi^{4}$.

Although there exists a possibility for generation of
a negative power low  potential in the models with dynamical
supersymmetry breaking (see, for example \cite{Binet}), such potential 
still looks to be exotic one in the context of the standard field theory.
As we see, in TMT such quintessential form of the effective potential
is obtained very easy and in a natural way.

Besides, adding any subleading (as $\phi\rightarrow\infty$) terms to 
(\ref{V12poz}) 
 does not alter the above results since their
relative contributions to $U(\phi)$ will be suppressed as $\phi$ is large
enough. In 
particular, adding  the 
term $V^{(0)}_{2}\int\sqrt{-g}d^{4}x$, $V^{(0)}_{2}\equiv const$,  which
in GR 
would have the sense of the cosmological constant term, does not affect
$U(\phi)$  as $\phi$ is large enough. Thus starting from the polynomial
form of the pre-potentials $V_1$ and $V_2$ with an appropriate choice
of the powers $n_1$ and $n_2$ of the leading terms, one can in fact provide a
generation of the inverse power low quintessential potential in such a way
that neither the cosmological constant problem nor the the problem of the 
flatness of the quintessential potential do not appear at all.

\subsection{ The 
exponential form of the TMT effective potential 
$U(\phi)$.}\label{subsec.exp}

A simple way to 
realize an exponential asymptotic form of the TMT effective potential 
$U(\phi)$, 
Eq.(\ref{VW}), is to define the pre-potentials $V_{1}$ and $V_{2}$ as 
follows: 
\begin{equation}
    V_{1}  =  
s_{1}m_{1}^{4}e^{\alpha \phi/M_{p}}; 
\qquad
V_{2}  = \frac{1}{4}
m_{2}^{4}e^{2\beta \phi/M_{p}}
\label{pot12} 
\end{equation}

Here $s_1 =\pm 1$ and we assume  that $\alpha$ and $\beta$ are positive 
constants.  The restrictions formulated after Eq.
(\ref{c})
have to be taken into account.
The effective TMT potential corresponding to the pre-potentials
(\ref{pot12}) 
\begin{equation}
    U=\frac{1}{m_{2}^{4}}\left( 
s_{1}m_{1}^{4}e^{-(\beta -\alpha)\phi/M_{p}}+ 
sM^{4}e^{-\beta \phi/M_{p}}\right)^{2}
\label{U1} 
\end{equation} 
contains two particular cases of special interest.
\medskip
\paragraph{The case $\alpha=\beta$}\label{par.expa=b}  
\medskip
This case corresponds to a sort of the scale invariant theory studied by 
Guendelman\cite{G}.  In fact, in this case the theory 
(\ref{totac}) is invariant under global transformations 
\begin{equation}
    g_{\mu\nu}\rightarrow 
e^{\theta}g_{\mu\nu}; \qquad
A_{\mu\nu\lambda}\rightarrow e^{\theta}A_{\mu\nu\lambda}
\label{st1} 
\end{equation} 
whereas the scalar field $\phi$ undergoes the 
shift
\begin{equation} 
    \phi\rightarrow \phi-\frac{M_{p}}{\beta}\theta 
\label{st3} 
\end{equation}
In such a model, the TMT effective potential has the form
\begin {equation}
U(\phi)=\frac{m_{1}^{8}}{m_{2}^{4}}\left[1+
\frac{s_{1}}{s}\left(\frac{M}{m_{1}}\right)^{4}
e^{-\beta \phi /M_{p}}\right]^{2}
\label{a=b}
\end{equation}
and the observation that $U(\phi)$ has an 
infinite flat region as $\phi\rightarrow\infty$ and approaches a 
nonzero constant $\frac{m_{1}^{8}}{m_{2}^{4}}$,  has been used in 
Ref. \cite{G} for discussion of possible cosmological 
applications with 
the choice  $\frac{s_{1}}{s}=-1$. The first possibility 
is related to the very early universe: a slow rolling (new inflationary)
scenario might be realized assuming that the universe starts at a 
sufficiently large value of $\phi$.  
Another scenario discussed by Guendelman in 
Ref.\cite{G} is based on a possibility for $\frac{m_{1}^{8}}{m_{2}^{4}}$
to be very small and this approach has the aim to construct a scenario for
the very late universe. 
In this 
scenario there could be a long lived stage with almost constant  energy 
density $\frac{m_{1}^{8}}{m_{2}^{4}}$ that will eventually disappear when
the universe achieves its true vacuum state with zero cosmological constant.
This occurs when 
the expression in parenthesis in Eq. (\ref{a=b}) becomes zero and therefore
no fine tuning is needed.
It turns out (see Refs. \cite{G}, \cite{G1}) that in
the presence of a matter, which is introduced in a way respecting the
global  symmetry (\ref{st1}), (\ref{st3}),  the change of the constraint 
(\ref{c}) leads  
to a correlation between $U(\phi)$ (close but not equal to zero) and the
 matter energy density.

In the case $\alpha =\beta$, the TMT effective potential 
(\ref{a=b}) is not a constant
due to the appearance of a non-zero integration constant $M$,
 that is actually, due to a spontaneous breaking of the global continuous 
symmetry
(\ref{st1}), (\ref{st3}). Guendelman noticed\cite{G} that in terms of the
dynamical variables used in the Einstein frame, that is 
$g^{\prime}_{\mu\nu}$ and $\phi$,  the symmetry transformations 
(\ref{st1}), (\ref{st3}) are reduced to  shifts (\ref{st3}) alone
($g^{\prime}_{\mu\nu}$ is invariant under transformations
(\ref{st1}), (\ref{st3})).
Thus in terms of the
dynamical variables of the Einstein frame, the spontaneous symmetry 
breaking is just that of the global continuous  symmetry 
$\phi\rightarrow \phi-\frac{M_{p}}{\beta}\theta$. It is important that,
as it was mentioned in Ref. \cite{G}, {\em this global continuous  
 symmetry is
restored as $\phi\rightarrow\infty$}. 

\medskip
\paragraph{The case $\beta
 >\alpha >0$.}\label{par.expb>a}
\medskip

This is the most interesting case from the viewpoint of the 
quintessence.
For $\beta \phi\gg M_{p}$,
the TMT 
effective potential (\ref{U1}) behaves as a decaying exponent:
\begin{equation}
    U \simeq \frac{m_{1}^{8}}{m_{2}^{4}} 
e^{-2(\beta -\alpha)\phi /M_{p}} \qquad as \qquad \beta \phi\gg M_{p}.
\label{U2}
\end{equation}
 
If we want to achieve the quintessential form of the TMT
effective potential (\ref{U2}) for not too large values of $\phi$
and with not too big difference in orders of $m_{1}$ and $M$ (this
point will be explained in the next section) then we need the 
condition 
\begin{equation}
0<\beta -\alpha\ll \beta
\label{fla<b}
\end{equation}
And of course the most evident argument in favour of this condition 
consists in the demand to provide
 the  flatness of the $\phi$-potential  at the late, $\phi$ -
dominated universe where it has to imitate the
present "cosmological constant". This is possible only if
 $\beta -\alpha$ is less or of order 1 while there are no reasons for 
$\beta$ not to be large in general.

Comparing this condition for $\alpha$ and $\beta$ with that of the 
model of Ref.
\cite{G} discussed just above, one can observe that 
the model
under consideration can be interpreted as that with a {\em small explicit
violation of the 
global symmetry} (\ref{st1}), (\ref{st3}). 
Notice that the expression for $U(\phi)$ as $\beta \phi\gg M_{p}$
does not include the integration constant $M$ and
the exponent is proportional to $\beta -\alpha$. This reflects the fact 
that such
asymptotic behavior of $U(\phi)$ results from the  explicit violation 
of the global continuous symmetry (\ref{st1}), (\ref{st3}). 

It is very interesting that although the discussed  {\em  global
continuous
 symmetry }
(\ref{st1}), (\ref{st3})  is broken in this model explicitly, the equtions
of motion show that the symmetry
is also restored   as 
$\phi\rightarrow\infty$, just as in the case $\alpha =\beta$ with only
spontaneous symmetry breaking.  Therefore, in terms of the
dynamical variables used in the Einstein frame, that is 
$g^{\prime}_{\mu\nu}$ and $\phi$, 
in the model where the condition (\ref{fla<b}) holds, the approximate 
global  symmetry 
$\phi\rightarrow \phi-\frac{M_{p}}{\beta}\theta$ is restored
as $\phi\rightarrow\infty$.

This observation opens unexpected
chance to solve the problem discussed by Carroll \cite{Carroll} (problem
5 in the list of problems in Introduction) which 
consists of
the following. There are no reasons to ignore a possibility that the
 scalar field $\phi$ interacts directly to
usual matter fields. Suppose that such interactions have the
form of the
coupling $f_{i}\frac{\phi}{m}{\cal L}_{i}$ where ${\cal L}_{i}$ is any
gauge invariant dimension-four operator, $m$ is a mass scale and
$f_{i}$ is a dimensionless coupling constant. The flatness of the
quintessential potential of the field $\phi$ means that excitations of
$\phi$ are almost massless. Therefore in the presence of 
direct interactions of
the $\phi$-field to the usual matter fields, one has to expect the
appearance of the very long-range forces which do not obey the
equivalence principle. Observational restrictions on such 
"fifth force"
impose  small upper limits on the coupling 
constants $f_{i}$. 

To explain smallness of $f_{i}$'s, Carroll  proposed to postulate 
that
the theory possesses an approximate global continuous 
symmetry of the form 
$\phi\rightarrow \phi+const$ (the idea similar to what is used in
 pseudo-Goldstone
boson models of quintessence\cite{Hill} where however, an explicit breaking
of the continuous chiral symmetry reduces it to a {\em discrete}
symmetry). In the framework 
of  Einstein's GR, such exact  continuous symmetry
is incompatible with a nontrivial potential of the scalar field $\phi$.
This means that if we were working in  Einstein's GR, then started 
from the model with the 
exact symmetry $\phi\rightarrow \phi+const$ and therefore with a 
{\em constant potential}, we would want to achieve a nontrivial, 
quintessential potential (passed also across a fine tuning purgatory)
  as a result of some mechanism for a symmetry
 breaking. Such a picture looks even more problematic one than 
the fine tuning problem itself. In addition, in the framework of such
general idea about a breaking of the symmetry 
$\phi\rightarrow \phi+const$, it is impossible to point out 
the parameters of the theory which could produce, after a symmetry
breaking, the small coupling constants $f_{i}$.

In contrast to GR, in TMT one can suppose that in yet unknown more
fundamental theory the global continuous symmetry (\ref{st1}), (\ref{st3}) 
is an exact one and $\alpha =\beta$. At energies below the Plank mass 
the symmetry is
breaking and it is assumed that the effective action describing the 
relevant physics, has the form of TMT, Eq. (\ref{totac})
 (inclusion of the usual matter will be studied
in Sec. VI),     with the
{\em nontrivial} pre-potentials (\ref{pot12}).
 The only thing we need 
from a mechanism for a symmetry breaking consists of  a small 
relative shift of the magnitudes
 of $\alpha$ and $\beta$ satisfying the condition (\ref{fla<b}).
If the symmetry breaking generates 
couplings of the scalar field $\phi$ to the usual matter fields, 
then 
the corresponding dimensionless coupling constants $f_{i}$ must be 
proportional\footnote{Notice that the exponents in the pre-potentials 
(\ref{pot12})
contain actually dimensional factors $\alpha M_{p}^{-1}$ and 
$\beta M_{p}^{-1}$.
Therefore the dimensionless parameter that could characterize the 
symmetry breaking has to be of the form $(\beta M_{p}^{-1}-
\alpha M_{p}^{-1})/\beta M_{p}^{-1}=(\beta -\alpha)/\beta$.}   
to some positive power of $(\beta -\alpha)/\beta$.

Notice that an unbounded increase of the pre-potentials as
$\phi\rightarrow\infty$ does not produce problems,
 at least on the classical level, since as was already mentioned
in Sec. II, the pre-potentials have no sense of a potential energy
density. The real potential is the TMT effective potential (\ref{VW})
that in the model under consideration approaches zero according to Eq.
(\ref{U2}) as $\phi\rightarrow\infty$. 

An evident generalization of the 
pre-potentials (\ref{pot12}) that maintains the behavior of $U(\phi)$
  as $\beta \phi\gg M_{p}$, Eq. (\ref{U2}),
consists of adding to them the terms with lower degree of
growth. They may be 
 for example, polynomials in 
$\phi$ as it was in the case discussed in previous subsection.
 Relative contributions 
of all adding terms into the TMT 
effective potential $U(\phi)$ will be  exponentially suppressed
for large  $\phi$.
If these additional terms 
appear as a result of  breaking of the symmetry (\ref{st1}), 
(\ref{st3}) (remined that $\alpha =\beta$ in the case of the exact symmetry),
 then  coefficients in front of them have to be proportional to 
some positive power
of the small parameter $(\beta -\alpha )/\beta $. 
The latter will be used in the next section.
For the same 
reasons as it was before,
the symmetry (\ref{st1}), (\ref{st3}) is restored as
$\phi\rightarrow\infty$.   

Simple reasoning adduced here as well as
in the previous subsection,
 does not look like trivial 
one if we recall that in GR adding
 any constant and/or increasing (as $\phi\rightarrow\infty$) term to the  
potential destined to be a quintessential one, causes a 
drastic violation of its desirable features: an arbitrary cosmological
constant appears and/or the flatness conditions are destroyed if no  
extreme fine 
tunings are made. The basis for the resolution of  these problems in TMT 
consists in a possibility to achieve a quintessence form of the effective 
potential
as $\phi$ is large enough, starting from pre-potentials {\em increasing}
as $\phi\rightarrow\infty$. As a matter of fact, this 
is the main advantage of the studied TMT models
 over the  quintessence models formulated in the framework of the
standard GR. 

In the conclusion it is worthwhile to notice for the following
discussion
that in all  cases considered 
in this section, $\chi^{-1}$ as 
the solution of the constraint (\ref{c}), asymptotically approaches zero
  as $\phi\rightarrow\infty$.

\section{Probe models: towards effective TMT potential \\
 of the "quintessential inflation" type}

\subsection{Some clarifications to the rest part of the paper}

The previous sections served a preparatory role in the formulation
and solution of the main problems of this paper.
In  Sec. III our attention was concentrated on the 
possibilities of TMT to generate without fine tuning
 the    scalar field $\phi$ potential which for $\phi$ large enough 
provides
a quintessence. It turns out however that some of such 
TMT effective potentials 
can appear to be also well defined to drive 
the 
early universe evolution.
In this paper I have no aim to look for a precise values of all 
{\em parameters} of the 
potential that
could be able to provide an adequate description of
 the cosmological evolution from slightly after
Planck time up to now and answer all demands of the realistic 
cosmology. But I do want to exhibit that the field theory models
 based on TMT provide the existence of a broad 
spectrum of tools giving us the firm belief that  such a potential 
can be generated without fine tuning.
More precisely, in this section I am going to demonstrate 
that exploring the results of the 
previous section  
one can make sure that TMT is able to
generate ({\em without any sort of fine tuning})   the 
effective potential of such a form that in the main could 
answer basic demands of  the realistic cosmology. 

Such qualitative examination is 
enough for the purposes of this paper which consist mainly in 
studying of some basic field-theoretic problems of TMT that turn out
to be in the very close interrelation with some fundamental features
of the cosmological scenario. The essence
of the matter is that generally speaking  the price for the success 
of TMT in the 
resolution of the cosmological constant problem is serious enough.
In fact, in order to incorporate the matter fields into the 
simplifying picture reviewed in Sec. II in such a way that the TMT
effective equations of motion of all fields in the Einstein frame 
would have the form of 
the  equations of motion of the standard field theory based 
on GR,
in Ref\cite{GK4}
we were forced to start from the very nonlinear (in the matter fields) 
original TMT action.
This circumstance makes the quantization of TMT a very hard problem
even on a semiclassical
level (i.e. the matter fields quantization
in the background curved space-time).  
We will see below
that for the second class cosmological scenarios (see the end of Sec. II)
with the appropriate choice of the pre-potentials, 
it is enough to start from the original TMT action with  
exactly the same degree of nonlinearity in matter fields
as in the standard theory in order to achieve the standard matter 
field theory in a background (pseudo-Riemannian) space-time. 
Then the matter fields quantization 
reduces  to the  
standard procedure of the matter fields quantization in curved 
 space-time\cite{Birrel}. Fortunately, it turns out
 that the 
choice of the initial 
cosmological conditions and the pre-potentials needed to provide such 
successful construction 
of the  matter field theory in the context of TMT,
corresponds to the class of models  where the TMT effective potential
allows to solve all five problems mentioned in Introduction.

\subsection{Models based on the hypothesis that the theory possesses
the explicitly broken global symmetry (\ref{st1}), (\ref{st3})}

The pre-potentials of the  form (\ref{pot12}) with additional 
(subleading as $\phi\rightarrow\infty$) terms
provide the possibility to generate 
the TMT effective potential $U(\phi)$ with an asymptotic 
 quintessence behavior that mimics the current
effective cosmological constant. For this to be done there is no 
need of any sort of fine tuning and the enough condition for this is  
 $0<\beta - \alpha\ll \beta$ in (\ref{pot12}).
 If however, one wants to extend the range of applicability
of the TMT effective potential of the same single scalar
field $\phi$ to satisfy constraints of the realistic cosmology
from inflation of the early universe up to the present-day universe,
 then we have too big arbitrariness in the choice of the 
additional terms to  (\ref{pot12}).
I restrict myself by models based on the 
idea that the action (\ref{totac})
 is the effective one
of a more fundamental theory  at
the energy scales below the Planck mass. It seems then to be natural
to suppose that transition from the fundamental theory to the 
effective one is accompanied by  
breaking of some fundamental symmetries. I will assume that one of 
such symmetries is the global one (\ref{st1}), (\ref{st3}) \footnote
{Of course, without knowledge of the fundamental theory one can not
discuss a mechanism for the symmetry breaking.}. Such approach
to the choice of prepotentials enables to narrow the amount of the
suitable versions. In particular, for models leading to the 
asymptotic (as $\phi\rightarrow\infty$) inverse
power low TMT effective potentials (discussed in Sec. III.A),
one can not point out a range where the symmetry (\ref{st1}), 
(\ref{st3}) is restored. This is why I am obliged to restrict 
myself by studying models of the type discussed in subsection III.B
and more precisely, by models where the condition (\ref{fla<b})
holds.

Below we will formulate three  
models where the modifications of the pre-potentials 
(\ref{pot12}) will be realized by adding the simplest terms 
explicitly breaking the symmetry (\ref{st1}), 
(\ref{st3}). 
The Planck mass $M_{p}$ is chosen as the 
typical scale for parameters of the dimension of mass  corresponding 
to the limit where the global symmetry (\ref{st1}), (\ref{st3}) is 
unbroken. Then the appearance of the mass 
parameters smaller than 
$M_{p}$ is a manifestation of a symmetry breaking by the 
appropriate terms since those 
parameters can be represented as $\left(\frac{\beta -
\alpha}{\beta}\right)^{n}M_{p}$, \quad $n>0$. In the framework of 
such an approach one can maintain that the model is {\em free of a fine
tuning} if orders of all such mass parameters are not too much differ 
from $M_{p}$ (in this connection see also discussions after Eqs. 
(\ref{U2}), (\ref{fla<b})
and footnote 1)   

\subsubsection{\bf Model 1}

\begin{equation}
V_{1}(\phi) =  m_{1}^{4}e^{\alpha \phi /M_{p}}; \qquad
V_{2}(\phi)  = \frac{1}{4}\left(4V_{2}^{(0)} +
m_{2}^{4}e^{2\beta \phi/M_{p}}
\right)
\label{model1}
\end{equation}

With the choice of the parameters $m_{2}=M_{p}$,\quad 
$4V_{2}^{(0)} =(10^{-3}M_{p})^{4}$, \quad 
$m_{1}=10^{-2}M_{p}$, 
 \quad $\beta =7$,\quad 
$\alpha =6$,
and with the integration constant $M^{4}=( 3q\times 
10^{-2}M_{p})^{4}$, \qquad $0<q \lesssim 1$ \quad($s=+1$), the TMT 
effective potential
$U(\phi)$, Eq. (\ref{VW}), is a monotonically decreasing function with the 
shape that
is convenient to describe in a piecewise form with the following four  
typical regions:

\begin{eqnarray}
U(\phi)&\approx & q^{8}M_{p}^4 \qquad for \qquad \phi <-2.2M_{p} ,
                          \nonumber\\
       &\approx & \frac{q^{8}M_{p}^{4}}{1+10^{12}e^{14\phi /M_{p}}}
                  \qquad for \qquad -2.2M_{p}<\phi < -1.8M_{p}
                          \nonumber\\
       &\approx &  10^{-12}M_{p}^{4}e^{-14\phi /M_{p}} \qquad for \qquad
                            -1.8M_{p}<\phi < 0.6M_{p} 
                          \nonumber\\     
       &\approx &  10^{-16}M_{p}^{4}e^{-2\phi /M_{p}} 
                    \qquad for \qquad \phi > 1.2M_{p}
 \label{mod1}
\end{eqnarray}

\subsubsection{\bf Model 2}

\begin{equation}
V_{1}(\phi)  =  \frac{1}{2}\mu_{1}^{2}\phi^{2}
+m_{1}^{4}e^{\alpha\phi /M_{p}}; \qquad
V_{2}(\phi)  =  \frac{1}{4}\left(4V_{2}^{(0)} 
+m_{2}^{4}e^{2\beta\phi /M_{p}}
\right)
\label{model2}
\end{equation}

With the choice of the parameters $m_2 =M_{p}$,\quad
$4V_{2}^{(0)} =(\frac{1}{3}M_{p})^4$, \quad
$\mu_{1} =10^{-4}M_{p}$, \quad $m_{1}=10^{-3}M_{p}$,
\quad  $\beta =7$,\quad 
$\alpha =6$ \quad
and with the integration constant $M^4= 
(\frac{1}{\sqrt{3}}10^{-2}M_{p})^4$,
\quad ($s=+1$), 
the TMT effective potential
$U(\phi)$, Eq. (\ref{VW}),  is a monotonically decreasing function with the shape that
one can describe in a piecewise form with the following three typical
 regions:

\begin{eqnarray}
U & \approx & \frac{1}{4}\lambda\phi^{4}, \qquad \lambda =10^{-14}
              \qquad for \qquad \phi < -\frac{1}{3}M_{p}
              \nonumber\\
  & \approx & 10^{-16}M_{p}^{4}\left[10^{-1}+
              \frac{1}{2}\left(\frac{\phi}{M_{p}}\right)^{2}\right]^{2}
              e^{-14\phi /M_{p}}
              \qquad for \qquad 0<\phi < 1.1M_{p} 
              \nonumber\\
  & \approx & 6\times 10^{-28}M_{p}^{4}e^{-2\phi /M_{p}}
              \qquad for \qquad  \phi > 1.4M_{p} 
\label{mod2}
\end{eqnarray}

where in the interval $0<\phi < 1.1M_{p}$ the factor in front of the 
exponential function varies very slowly.

\bigskip

\subsubsection{\bf Model 3}

\begin{equation}
V_{1}(\phi)  =  \frac{1}{2}\mu_{1}^{2}\phi^{2}
+m_{1}^{4}e^{\alpha \phi/M_{p}}; \qquad
V_{2}(\phi)  =  \frac{1}{4}\left(4V_{2}^{(0)} 
+\frac{1}{2}\mu_{2}^{2}\phi^{2}
+m_{2}^{4}e^{2\beta \phi/M_{p}}
\right)
\label{model3}
\end{equation}

With the choice of the parameters $m_2 =M_{p}$,\quad
$4V_{2}^{(0)} =(10^{-1}M_{p})^4$, \quad
$\mu_{1}=10^{-4}M_{p}$,\quad
$m_{1} = 10^{-3}M_{p}$,\quad $\mu_{2}=10^{-2}M_{p}$,\quad
$\beta =7$ , \quad
$\alpha =6$
and with the integration constant 
$M^4=(\frac{1}{\sqrt{3}}10^{-2}M_{p})^{4}$,
\quad ($s=+1$), 
the TMT effective potential, Eq. (\ref{VW}), 
is a monotonically decreasing function with the shape that
one can describe in a piecewise form with the following three typical
 regions:

\begin{eqnarray}
U & \approx & \frac{1}{2}m^{2}\phi^{2}, \qquad m=10^{-6}M_{p}
               \qquad for \qquad  \phi < -0.7M_{p}
               \nonumber\\
  & \approx & 10^{-16}M_{p}^{4}\left[10^{-1}+
         \frac{1}{2}\left(\frac{\phi}{M_{p}}\right)^{2}\right]^{2}
         e^{-14\phi /M_{p}}
         \qquad for \qquad -0.6<\phi < 1.5M_{p}
               \nonumber\\ 
  & \approx & 10^{-24}M_{p}^{4}e^{-2\phi /M_{p}}
         \qquad for \qquad \phi > 1.7M_{p}
\label{mod3}
\end{eqnarray}
where in the interval $-0.6M_{p}<\phi < 1.5M_{p}$
the factor in front of exponential function  varies very slowly.

\bigskip

\subsection{\bf Some general features of the models 1 - 3.}

As it was already noted, the exact fitting of all parameters to 
satisfy the requirements of the realistic cosmology is over and 
above the plan of this paper. Our aim here is rather a demonstration 
of 
extremely broad spectrum of tools giving by TMT to solve some 
fundamental problems
of the realistic cosmology. 

1) In each of the models 1-3 with the action  (\ref{totac}), 
the global continuous symmetry (\ref{st1}), (\ref{st3}) 
is violated  by all terms of $V_1$  and $V_2$ except for the last
term of $V_2$. {\em The symmetry is 
restored at the limit} $\phi\rightarrow\infty$.   
All mass parameters (including mass parameters corresponding to 
"$\Lambda$-terms" in each of the models) have orders equal or
slightly less than the Planck
mass (but not less than the GUT  scale). 

2) One can see that the TMT effective potential $U(\phi)$ of each of 
the models 
1 - 3 has a region that can be responsible for an inflation
of the early universe. 
Let us refer to this  region of  $U(\phi)$ as the {\em "inflationary
region"} of $U$. 

In  model 1 the  inflationary region of $U$ is  the infinite
interval $-\infty < \phi <-1.8M_{p}$  with practically constant value
$U(\phi)\approx q^{8}M_{p}^4$ that smoothly passes on a slowly decreasing 
region. Such inflationary region of $U$ might be responsible for
an initial stage of a new inflationary scenario\cite{Linde-Stein}.

In models 2 and 3, the  inflationary regions of $U$  
have the form of the power low potentials ($\frac{1}{4}\lambda \phi^4$
 and $\frac{1}{2}m^2 \phi^2$ respectively)  driving the chaotic 
inflation \cite{Linde1}. Parameters of the pre-potentials are chosen
in such a way that the inflationary region of $U$ satisfies the 
requirements of the realistic cosmology. It is very 
important to stress that this can be done without  strong tuning 
of the parameters, in contrast with the  GR approach 
to the chaotic inflation models where the  
strong enough tuning is needed. The choice of $\beta$ and $\alpha$ does 
not affect practically
the inflationary region of $U(\phi)$.   
 
3) The TMT effective potential $U(\phi)$ of each of the 
models 1 - 3 behaves as 
\begin{equation}
U(\phi)\approx \frac{m_{1}^{8}}{M_{p}^{4}} e^{-2(\beta-\alpha)\phi/M_{p}}
\qquad as\qquad \phi > \phi_{b} =\varpi M_{p},
\label{quint}
\end{equation}
where the constant factor $\varpi$  of order 1  is very sensitive
to the choice of parameters. Let us refer to this  region of  $U(\phi)$ as
 the {\em "quintessential 
region"} since it can serve for the 
quintessential model of the present universe. One should  make 
here an important remark. The quintessential 
region of $U$ has the form (\ref{quint}) where the  value of 
$(\beta -\alpha)/\beta \ll \beta$ 
determines a strength of the symmetry breaking. The choice of
$\beta -\alpha =1$ and $\beta =7$ in the models 1-3 has just an 
illustrative aim and
it is not a problem to adjust the  value of $\beta -\alpha$ to satisfy
the observable value of $\Omega_{\phi}$ at present.

4) Between the  inflationary and quintessential regions,  
there exists 
an {\em intermediate region} of great interest. 
The TMT effective potential $U(\phi)$ in the intermediate region 
 can be represented in the general form
\begin{equation}  
U(\phi)=f(\phi)M_{p}^{4}
e^{-2\beta\phi/M_{p}} 
\label{inter}
\end{equation}
where $f(\phi)$ is a very slowly varying function compared to the 
exponential factor. There is a 
remarkable property 
of the intermediate region of $U$ that provides possibilities for 
resolution of some fundamental problems
of the realistic cosmology: by an appropriate
choice of $\beta$ one can achieve a very rapid decreasing of 
$U(\phi )$  after inflationary epoch that provides conditions for transition 
to the radiation and matter dominated era. 
For instance, in model 3 the TMT effective potential $U(\phi)$ 
at the end of the intermediate region ($\phi\approx 1.5M_{p}$) is
roughly $10^{28}$ times less than at the beginning 
of the intermediate region ($\phi\approx -0.6M_{p}$). This property 
of the intermediate region of $U$ may be very useful for resolution of
 problems of cosmological nucleosynthesis constraints
 and large-scale structure formation\cite{Wett}, 
\cite{Ferr-Joyce}, \cite{Birkel}, \cite{Stein}. 
 The exact shape of the intermediate region of $U$ (steepness and 
the range of definition)
dictated by the realistic cosmology can be adjusted by the choice of
the magnitudes of $\alpha$, $\beta$ and dimensional parameters 
(like $M$, $\mu_{1}$
etc.).  

5) 
Combining the intermediate and the quintessential regions one can
see that the {\em post-inflation region} of the TMT effective potential 
can be
represented approximately in the exponential form described by 
Eq.(\ref{expintrod}) 
 {\em with 
$\phi$-dependent parameter} $a$ (see Ref. \cite{Wett})
\begin{eqnarray}
a=a(\phi) & = & 2\beta \qquad as\qquad \phi < \phi_{b}
                       \nonumber\\
          & = & 2(\beta -\alpha) \qquad as \qquad\phi > \phi_{b}
\label{vara}                       
\end{eqnarray}
where $\phi_{b}$ (see Eq. (\ref{quint})) is a boundary value of $\phi$ 
between the 
intermediate and quintessential regions of 
$U(\phi)$. It seems to be very attractive that this result is obtained
in a natural way in the framework of the field theory model
without any assumptions specially intended for this. The only thing 
have been assumed is that the model possesses the approximate global 
continuous symmetry (\ref{st1}), (\ref{st3}) and the  value of 
$\beta -\alpha \ll \beta$ 
depending on a strength of the symmetry breaking should not be large
in order to provide the flatness  of the TMT effective
potential in the quintessential region. 

6) It turns out that in the models 2 and 3
the shape of $U(\phi)$ in the buffer range between the 
inflationary 
and  intermediate regions  can be very sensitive to 
 variations of the parameters entering into $V_1$ and $V_2$.
By means of a suitable change of the parameters one can achieve (without
altering the qualitative properties of the discussed above regions), 
for instance
an almost flat shape of $U$ in this buffer range or even successive 
local minimum and 
maximum immediately after the inflationary region. 
This feature of 
the models may be very important  if for example one wants to realize 
a scenario where  the instant preheating\cite{Linde2} occurs before entry
into the intermediate region.  

The final remarks concerns the terminology. Since the scalar field
$\phi$, in context of models 1-3, dominates both at the very early and
at the late universe, acting in such a way that the universe expands with
acceleration, let us call it  the inflaton field following the terminology
by Peebles and Vilenkin\cite{PV}.

\section{Inclusion of usual matter fields}

\subsection{Outline of the approach to the problem}

Inclusion of the ordinary 
matter fields (like
vector bosons, fermions, etc.) in TMT  is a very  nontrivial problem. 
In the framework
of the first class cosmological scenarios, there was
shown in Ref. \cite{GK4} that the field theory model exists where in the
conformal Einstein frame,
the classical equations of motion of the gauge unified theories as well as
the GR equations  are exactly reproduced. The merit of this model is
that the spontaneous symmetry breaking (SSB) does not generate the cosmological
constant term. However a serious defect of this model consists of the
necessity to use the artificial enough form of how the  gauge fields
kinetic terms and the
fermions selfinteractions enter into the original action. This creates
a situation where it is unclear how one can approach  the
matter fields quantization. 

The  origin of the problem is practically reduced to the role of the 
constraint (\ref{c}) which is modified in the presence of usual matter 
fields. In fact,  matter fields in general contribute 
to the constraint and  then the $\chi$-field becomes depending of 
matter fields. Therefore when starting with
Lagrangians $L_{1}$ and $L_{2}$ including the matter fields 
in a form similar to  the canonical one, the resulting matter fields 
equations of motion in the Einstein picture (obtained with 
the use of the conformal transformations (\ref{ct}) or their generalization 
in  the presence of fermions) can appear in general to be very nonlinear. 

Inclusion of the usual matter fields in the context of
the models of Sec. IV.B permits to avoid this problem. In fact, 
following the idea that the only  mass scale typical for the inflaton
physics in the limit where the symmetry (\ref{st1}), (\ref{st3}) is exact,
 is the Planck mass, and terms that explicitly breaks this 
symmetry, contain mass parameters only a few orders of magnitude less than
 $M_{p}$, we provide a situation where {\em the usual matter fields 
contributions
to the constraint  appear to be negligible in 
comparison with the
inflaton  contributions throughout the history of the universe}. At the late
universe, the unbounded increase of the pre-potentials (as 
$\phi\rightarrow\infty$) reinforces this effect.
As a result of this, the scalar field $\chi$ with high accuracy is determined
by the same constraint (\ref{c}) as it was in the absence of the usual 
matter fields. This allows,  starting from the Lagrangians similar to 
usual ones, one to keep after transition to the Einstein frame
the desirable basic features of the usual matter fields sector. 
Together with the basic idea about the broken continuous
global symmetry (\ref{st1}), (\ref{st3}) modified to the case of the 
presence of fermions, this approach provides  possibilities for 
constructing  gauge models in the context of TMT and, at the same
time, to solve problems 1-5 of Introduction.

\subsection{Action of a gauge abelian model and continuous global symmetry }

In the framework of the formulated above general ideas let us consider
a toy model that possesses gauge abelian symmetry and contains the 
following matter fields: a complex scalar
field $\xi =\frac{1}{\sqrt{2}}(\xi_{1}+i\xi_{2})$, an abelian gauge vector 
field $A_{\mu}$ and a fermion $\Psi$.
Generalization to non-abelian
 gauge  theories can be performed straightforward.

In the presence of fermions, the vierbein-spin-connection formalism
\cite{Gasp}, \cite{Freund}
 has
to be used instead of the first order formalism of Sec. II. The action of
the model has the general form as in Eq. (\ref{S}) with
\begin{eqnarray}
L_{1}=-\frac{1}{\kappa}R(\omega ,V)+
\frac{1}{2}g^{\mu\nu}\phi,_{\mu}\phi,_{\nu}
+ g^{\mu\nu}(\partial_{\mu}
-ieA_{\mu})\xi (\partial_{\nu}+ieA_{\nu})\xi^{*}-
V_{1}(\phi ,|\xi |)+
\nonumber\\
+\frac{i}{2}\overline{\Psi}\left\{\gamma^{a}V_{a}^{\mu}
(\overrightarrow{\partial}_{\mu}+\frac{1}{2}\omega_{\mu}^{cd}\sigma_{cd}-
ieA_{\mu})
-(\overleftarrow{\partial}_{\mu}-\frac{1}{2}\omega_{\mu}^{cd}\sigma_{cd}+
ieA_{\mu})
\gamma^{a}V_{a}^{\mu}\right\}\Psi
 \label{L1gr+mat}
 \end{eqnarray}
\begin{equation}
L_{2}=V_{2}(\phi)-
\frac{1}{4}g^{\alpha\beta}g^{\mu\nu}F_{\alpha\mu}F_{\beta\nu}-
h\overline{\Psi}\Psi|\xi | e^{\gamma\phi/M_{p}}
 \label{L2gr+mat}
 \end{equation}

Here the following definitions are used\cite{Gasp}:
\begin{equation} 
R(\omega ,V) =V^{a\mu}V^{b\nu}R_{\mu\nu ab}(\omega); \quad
R_{\mu\nu ab}(\omega)=\partial_{\mu}\omega_{\nu ab}
+\omega_{\mu a}^{c}\omega_{\nu cb}
-(\mu\leftrightarrow\nu).
        \label{B}
\end{equation}
where $V^{a\mu}=\eta^{ab}V_{b}^{\mu}$, $\eta^{ab}$ is the diagonal
$4\times 4$
matrix with elements $(1, -1,-1,-1)$ on the diagonal, $V^{a}_{\mu}$
are the vierbeins 
and $\omega_{\mu}^{ab}=-\omega_{\mu}^{ba}$ 
($a,b=0,1,2,3$) is
the spin connection.

Pre-potential $V_{2}(\phi)$ is the same as in the 
models of Sec. IV.B. Pre-potential $V_{1}(\phi ,|\xi |)$  is
chosen in the form
\begin{equation}
V_{1}(\phi ,|\xi |)= V_{1}(\phi)+P(|\xi |)e^{\alpha\phi/M_{p}}
\label{potV}
\end{equation}
where $V_{1}(\phi)$ is the same as in the models of Sec. IV.B, that is 
$e^{\alpha\phi/M_{p}}$ is the common factor in front of 
$m_{1}^{4}+P(|\xi |)$ in (\ref{potV}).  

The transformations of the  continuous
global symmetry (\ref{st1}), (\ref{st3}) are generalized now to the form
\cite{G1}
\begin{eqnarray}
V^{\mu}_{a}\rightarrow e^{-\theta /2}V^{\mu}_{a};\qquad    
g_{\mu\nu}\rightarrow 
e^{\theta}g_{\mu\nu}; \qquad
A_{\mu\nu\lambda}\rightarrow e^{\theta}A_{\mu\nu\lambda}; \qquad
    \phi\rightarrow \phi-\frac{M_{p}}{\beta}\theta
\nonumber\\
 \xi\rightarrow \xi ;\qquad A_{\mu}\rightarrow A_{\mu}; \qquad
\Psi \rightarrow e^{-\theta /4}\Psi ;\qquad
\overline{\Psi} \rightarrow e^{-\theta /4} \overline{\Psi} 
\label{fullglobal} 
\end{eqnarray}

The term $\int P(|\xi|)e^{\alpha\phi/M_{p}}\Phi d^{4}x$ breaks the symmetry
(\ref{fullglobal}) by the same manner as  the pre-potential
$V_{1}(\phi)$.
For the "Yukawa coupling type" term
\begin{equation} 
S_{Yuk}=-h\int \overline{\Psi}\Psi |\xi | e^{\gamma\phi/M_{p}}
\sqrt{-g} d^{4}x
\label{SYuk, original}
\end{equation}
to be  invariant under transformations (\ref{fullglobal}), the parameter
 $\gamma$ must be $\gamma =\frac{3}{2}\beta$. The  value of 
$\gamma$ preferable from the dynamical point of view will be discussed 
later and we will see that $\gamma <2\beta$.
 All other
terms describing the usual matter  fields are invariant under transformations
(\ref{fullglobal}). If $\gamma \neq\frac{3}{2}\beta$ then the symmetry is
explicitly broken only by the Yukawa coupling type term and by 
pre-potentials $V_{1}(\phi ,|\xi |)$ and $V_{2}(\phi)$.
Thus, similar to the models of Sec. IV.B, in the model
with the Lagrangian densities (\ref{L1gr+mat}) and (\ref{L2gr+mat}),
 the global continuous 
symmetry (\ref{fullglobal} ) is restored as $\phi\rightarrow\infty$.

It is interesting that 
the form of the $\phi$-dependence of the "Yukawa type" term 
dictated by the symmetry (\ref{fullglobal}) is very similar to a motivated
by string theories
"nucleon-scalar coupling" discussed by Wetterich\cite{Wett} in the
context of a quintessence type model with exponential potential.

Note finally that for "pedagogical" reason we have started from the 
simplified model where 
the Yukawa type term appears only with the measure $\sqrt{-g}$. We will
see later (see Sec. VI.H) that an additional Yukawa type term 
in (\ref{L1gr+mat}), that is with the measure $\Phi$, is needed 
to provide a possibility to avoid the long-range force problem.  

\subsection{Connection, equations of motion and constraint}

Variation of the action  with respect to 
$\omega_{\mu}^{ab}$ leads to the equation solution of which
is represented in the form\cite{GK4}
\begin{equation}
\omega_{\mu}^{ab}=\omega_{\mu}^{ab}(V) + K_{\mu}^{ab}(\sigma)  + 
K_{\mu}^{ab}(V,\overline{\Psi},\Psi)
\label{C4}
 \end{equation}
where $\omega_{\mu}^{ab}(V)$
is the Riemannian part of the connection\cite{Gasp}, \cite{Freund}
and
\begin{equation}
K_{\mu}^{ab}(\sigma)=\frac{1}{2}\sigma_{,\alpha}(V_{\mu}^{a}V^{b\alpha}-
V_{\mu}^{b}V^{a\alpha}), \qquad \sigma \equiv\ln\chi ,
\label{C6}
 \end{equation}
\begin{equation}
K_{\mu}^{ab}(V,\overline{\Psi},\Psi)=
\frac{\kappa}{8}\eta_{ci}V_{d\mu}\varepsilon^{abcd}\overline{\Psi}
\gamma^{5}\gamma^{i}\Psi.
\label{C7}
 \end{equation}

For short we omit here equations obtained by variations of vierbeins,
$A_{\mu\nu\lambda}$ as well as of the matter fields
$\phi$, $\xi$, $A_{\mu}$, $\Psi$ and 
$\overline{\Psi}$. 
Combining equations obtained by variation of vierbeins and
$A_{\mu\nu\lambda}$ and using equations of motion for 
$\Psi$ and $\overline{\Psi}$,  
one can eliminate $R(\omega ,V)$ and the result is the constraint
\begin{equation}
sM^{4}+V_{1}(\phi)+P(\varphi )e^{\alpha\phi /M_{p}}
=\frac{2}{\chi}\left[V_{2}(\phi)-
\frac{3}{4\sqrt{2}}h\overline{\Psi}\Psi \varphi e^{\gamma\phi/M_{p}}\right]
\label{con+matt}
\end{equation}
which is a direct generalization of the constraint (\ref{c}) to the model
we study here.

One of the aims of this toy model 
consists in a demonstration of a possibility to construct  realistic gauge 
unified
theories (like electro-weak and GUT) in the context of cosmological
scenarios dictated by models of Sec.
IV.B. Introducing the scalar field $\xi$ is intended for realization of the
Higgs phenomenon. 
Since
$P(\varphi)$ and $m_{1}^{4}$ appear in the combination 
$m_{1}^{4}+P(\varphi)$,   the constant 
part of $P(\varphi)$ can be always absorbed by $m_{1}^{4}$. Then  
it is natural to assume\footnote{Recall that  $m_{1}$ appears in the 
definition of the pre-potential $V_{1}(\phi)$ in models of Sec. IV.B,
and the values of $m_{1}$ are chosen such that
$m_{1}^{4}=(10^{-2}M_{p})^{4}$ in the model 1 and  
$m_{1}^{4}=(10^{-3}M_{p})^{4}$ in models 2 and 3.} 
 that $|P(\varphi)|\ll m_{1}^{4}$. Later, 
turning to quantum effective potential, 
we will discuss a concrete model where $P(\varphi)=
\frac{\overline{\lambda}}{4!}\varphi^{4}$ and then the idea explained in 
Sec. VI.A becomes more clear : the choice of 
the mass parameters in the models of Sec. IV.B allows to provide a 
situation where the contribution of the Higgs field $\varphi$
to the constraint (\ref{con+matt}) 
is negligible with respect to the inflaton field $\phi$-contribution
 and hence it can give only extremely small corrections to
the main picture .
 If fluctuations of  fermionic fields are not anomaly large, it 
is natural to expect 
that the same conclusion  is  true 
for fermionic contribution to the constraint (\ref{con+matt}) as well.
So, the $\chi$-field determined by the constraint 
(\ref{con+matt}), in  practically interesting cases
coincides with the $\chi$-field determined by the 
constraint (\ref{c}) which holds in the model free of the usual matter 
at all. 
For short, in what follows,
when neglecting the usual matter fields contribution to the constraint,
 we will use the term "A-approximation".
This notion will be very useful in the next subsection where
 we are going to represent equations of motion
in the Einstein frame.

\subsection{Equations of motion for the selfconsistent problem in the
Einstein frame}

In the presence of fermions, the transition to the "Einstein frame" 
(more suitable term for this case would be the Einstein-Cartan frame) is
carried out by the transformations to the new variables\cite{GK4}
\begin{eqnarray}
V_{a\mu}(x)\rightarrow V^{\prime}_{a\mu}(x)=\chi^{1/2}(x)V_{a\mu}(x); 
\qquad 
g_{\mu\nu}(x)\rightarrow g^{\prime}_{\mu\nu}(x)=\chi (x)g_{\mu\nu}(x);
 \nonumber\\
\Psi (x)\rightarrow\Psi^{\prime} (x)=\chi^{-1/4}(x)\Psi (x); \qquad 
\overline{\Psi}(x)\rightarrow\overline{\Psi}^{\prime}(x)=
\chi^{-1/4}(x)\overline{\Psi}(x);
 \nonumber\\
\phi\rightarrow\phi ; \qquad 
A_{\mu\nu\lambda}\rightarrow A_{\mu\nu\lambda};\qquad
\varphi\rightarrow\varphi ; \qquad A_{\mu}\rightarrow A_{\mu},
 \label{ctferm}
\end{eqnarray}
where $\chi$ is determined by the constraint (\ref{con+matt}).

In fact, after transition to the new variables defined by the 
transformations (\ref{ctferm}), the $\sigma$-contribution 
(\ref{C6}) to the 
spin connection is
canceled and the transformed spin connection takes the form\cite{GK4}
\begin{equation}
\omega_{\mu}^{\prime cd}=\omega_{\mu}^{cd}(V^{\prime})+
\frac{\kappa}{8}\eta_{ci}V^{\prime}_{d\mu}\varepsilon^{abcd}
\overline{\Psi}^{\prime}\gamma^{5}\gamma^{i}\Psi^{\prime}.
 \label{connEin}
 \end{equation}
that coincides with the well-known solution for the spin 
connection in the context of the first order formalism approach
to the Einstein-Cartan theory\cite{Gasp} where a Dirac
spinor field is the only source of a non-riemannian part of the 
connection. Hence the curvature tensor (\ref{B}) expressed in terms of
the new connection (\ref{connEin}) becomes the curvature tensor of such
an Einstein-Cartan theory.\footnote{Notice that in the original frame, the
terms including $\sigma_{,\mu}$ (recall that $\sigma\equiv\ln\chi$) 
originate
a non-metricity and therefore TMT in the original variables has
no form of an Einstein-Cartan theory.} 

At the same time, in the 
 fermionic field equation , all terms containing  
$\sigma_{,\mu}$ also disappear\cite{GK4} in the 
Einstein-Cartan
 frame and the result is
\begin{equation}
\left\{i\left [V_{a}^{\prime\mu}\gamma^{a}\left
(\partial_{\mu}-ieA_{\mu}\right )+
\gamma^{a}C^{\prime b}_{ab}
+\frac{i}{4}\omega_{\mu}^{\prime cd}\varepsilon_{abcd}\gamma^{5}\gamma^{b}
V^{\prime a\mu}\right ]
-\frac{h}{\sqrt{2}}\varphi \frac{e^{\gamma\phi/M_{p}}}{\chi^{3/2}}
\right\}\Psi^{\prime}=0
 \label{PsiEin}
 \end{equation}
where $C^{\prime b}_{ab}$ 
is the trace of the Ricci rotation coefficients\cite{Gasp} in the new variables
and the unitary gauge is used: after a shift we define
\begin{equation}
\xi =\frac{1}{\sqrt{2}}\varphi\equiv\frac{1}{\sqrt{2}}(\upsilon +
\tilde{\varphi }(x)); \qquad \upsilon =const
\label{varphidefinition}
\end{equation}
Equation for $\overline{\Psi}^{\prime}$ has  similar structure. 
The only 
difference of these fermionic equations from the standard 
Dirac equations in the Einstein-Cartan theory\cite{Gasp}  
is related to an unusual
  Yukawa type term and it will be discussed later. Notice
that for purposes of realistic particle physics one can neglect the second 
term in Eq. (\ref{connEin}) that leads to a "spin-spin contact interaction"
\cite{Gasp}
with  coupling constant $M_{p}^{-2}$. For short, in what follows,
when neglecting this interaction, we will use the term "B-approximation".

Other equations of motion in the Einstein-Cartan frame have the 
following form:
\begin{equation}
\frac{1}{\sqrt{-g^{\prime}}}\partial_{\mu}(\sqrt{-g^{\prime}}
g^{\prime\mu\nu}\partial_{\nu}\phi)+ 
\frac{1}{\chi}\left[\frac{dV_{1}}{d\phi}-\frac{1}{\chi}\frac{dV_{2}}{d\phi}
+\frac{\alpha}{M_{p}}P(\varphi)e^{\alpha\phi /M_{p}}\right]=
-\frac{h\gamma}{\sqrt{2}M_{p}}\overline{\Psi}^{\prime}\Psi^{\prime} 
\varphi \frac{e^{\gamma\phi/M_{p}}}{\chi^{3/2}}; 
\label{phiEin}
\end{equation}

\begin{equation}
\frac{1}{\sqrt{-g^{\prime}}}\partial_{\mu}(\sqrt{-g^{\prime}}
g^{\prime\mu\nu}\partial_{\nu}
\tilde{\varphi})+
\frac{e^{\alpha\phi /M_{p}}}{\chi}\frac{dP(\varphi)}{d\varphi}-
e^{2}\varphi g^{\prime\alpha\beta}A_{\alpha}A_{\beta}=
-\frac{h}{\sqrt{2}}
\overline{\Psi}^{\prime}\Psi^{\prime} 
\frac{e^{\gamma\phi/M_{p}}}{\chi^{3/2}}; 
 \label{varphiEin}
\end{equation}

\begin{equation}
\frac{1}{\sqrt{-g^{\prime}}}\partial_{\mu}\left(\sqrt{-g^{\prime}}
g^{\prime\mu\alpha}g^{\prime\nu\beta}
F_{\alpha\beta}\right)
+ \frac{e^{2}}{2}\varphi^{2}g^{\prime\mu\nu}A_{\mu}=
-e\overline{\Psi}^{\prime}\gamma^{a}V^{\prime\mu}_{a}\Psi^{\prime}
\label{AEin}
\end{equation}

It is very important to stress that in the A and B-approximations, all 
matter fields equations, 
(\ref{PsiEin}), 
(\ref{phiEin})-(\ref{AEin}), have the {\em canonical structure} of the 
corresponding matter fields equations in a Riemannian space-time. The 
only specific features of these equations are concentrated in unusual
forms of the effective potentials and  some  of the interactions .  

After some
algebraic manipulations with equations resulting from variation of the 
vierbeins,
transition to the new variables by means of (\ref{ctferm}) and
 making use the fermionic equation (\ref{PsiEin}) and similar equation
for $\overline{\Psi}^{\prime}$, we obtain  {\em canonical}
 gravitational
equations of the Einstein-Cartan theory. If  finally one to write down
these equations in the B-approximation,
we come to the {\em canonical GR gravitational equations }
\begin{equation}
G_{\mu\nu}=\frac{\kappa}{2}T_{\mu\nu}
 \label{EcE}
\end{equation}
where $G_{\mu\nu}$ is the Einstein tensor of the Riemannian space-time
with metric $g^{\prime}_{\mu\nu}$
and the energy-momentum tensor has a canonical GR structure\cite{Birrel}:
\begin{eqnarray}
 T_{\mu\nu} & = & \phi_{,\mu}\phi_{,\nu}-
\frac{1}{2}g^{\prime}_{\mu\nu}\phi_{,\alpha}\phi_{,\beta}
g^{\prime\alpha\beta}+\frac{1}{\chi^{2}}V_{2}(\phi)g^{\prime}_{\mu\nu}
 + \tilde{\varphi}_{,\mu}\tilde{\varphi}_{,\nu}-
\frac{1}{2}g^{\prime}_{\mu\nu}
\tilde{\varphi}_{,\alpha}\tilde{\varphi}_{,\beta}g^{\prime\alpha\beta}
\nonumber\\
 & + & \frac{1}{4}g^{\prime}_{\mu\nu}F_{\alpha\beta}F_{\tau\rho}
g^{\prime\alpha\tau}g^{\prime\beta\rho}
-F_{\mu\alpha}F_{\nu\beta}g^{\prime\alpha\beta}
 + e^{2}(\upsilon +\tilde{\varphi})^{2}\left(A_{\mu}A_{\nu}-
\frac{1}{2}g^{\prime}_{\mu\nu}A_{\alpha}A_{\beta}g^{\prime\alpha\beta}\right)
\nonumber\\
& + & \frac{i}{2}\left[\overline{\Psi}^{\prime}\gamma^{a}V_{a(\mu }^{\prime}
\nabla_{\nu )}\Psi^{\prime}-(\nabla_{(\mu}\overline{\Psi}^{\prime})
\gamma^{a}V_{\nu )a}^{\prime}\Psi^{\prime}\right] 
\label{Tmunu}
\end{eqnarray}
where $\nabla_{\mu}\Psi^{\prime} =\left(\partial_{\mu}+
\frac{1}{2}\omega_{\mu }^{\prime cd}\sigma_{cd}-
ieA_{\mu }\right)\Psi^{\prime}$
and $\nabla_{\mu}\overline{\Psi}^{\prime}=
\partial_{\mu}\overline{\Psi}^{\prime}-
\frac{1}{2}\omega_{\mu}^{\prime cd}\overline{\Psi}^{\prime}\sigma_{cd}+
ieA_{\mu}\overline{\Psi}^{\prime}$.

Notice again that $\chi$-field entering into Eqs. (\ref{PsiEin}),
(\ref{phiEin}), (\ref{varphiEin}) and(\ref{Tmunu}), is determined by
the constraint (\ref{con+matt}) which in the A-approximation gives
\begin{equation}
\frac{1}{\chi}=\frac{M^{4}+V_{1}(\phi)}{2V_{2}(\phi)}.
\label{1/chi}
\end{equation}

In what follows, all discussion will be 
performed in the framework of A- and B-approximation. 

It is worthwhile to notice  that the transformations of the global 
continuous
symmetry (\ref{fullglobal}) expressed in terms of the variables of 
the  Einstein frame, are reduced just to shifts of $\phi$:
$\phi\rightarrow\phi -\frac{M_{p}}{\beta}\theta$.

 \subsection{Effective classical action
for usual matter fields in the background}

To study  the matter fields sector of the system of equations
(\ref{PsiEin}) - (\ref{1/chi})  one 
has to define an appropriate background. In the models
where the usual matter was absent and the inflaton field $\phi$ was
 the only field of the non-gravitational sector, 
 the  gravitational background in the variables of the Einstein frame
 is described by
  external field $g^{\prime}_{\mu\nu}$.
 If however, we want to construct quantum theory of the usual matter fields
then it seems to be natural to start from the approximation where in the 
addition to
the  gravitational background, the inflaton field $\phi$ is also regarded as 
the background one. This can be done
since in a course of its  evolution, the classical inflaton field $\phi$
 remains practically constant 
during a typical time of quantum fluctuations of the matter fields. 

So, let us study some features of the particle physics model in the 
background
which, in terms of variables of the Einstein picture, consists of two 
external fields: $g^{\prime}_{\mu\nu}$ and $\phi$. For
short I will refer to this issue as the "particle physics model in the 
cosmological background". 

The effective classical action for the particle physics model 
corresponding to the system of equations 
(\ref{PsiEin}),
(\ref{varphiEin}) and (\ref{AEin}), 
in the cosmological background,  can be written down in the following form 
(in the unitary gauge)
\begin{eqnarray}
S_{class}^{background} & = &\int\sqrt{-g^{\prime}}\left[ 
 \frac{1}{2}g^{\prime\mu\nu}\varphi,_{\mu}\varphi,_{\nu}
-V_{cl}(\varphi \, ; \, \phi)+
\frac{e^{2}}{2}\varphi^{2}A_{\mu}A_{\nu}g^{\prime\mu\nu}\right.
\nonumber\\
 & - &\left.\frac{1}{4}
g^{\prime\alpha\beta}g^{\prime\mu\nu}F_{\alpha\mu}F_{\beta\nu}
+L_{kin}(\overline{\Psi^{\prime}}, \Psi^{\prime}, A_{\mu})
+L_{Yuk}(\overline{\Psi^{\prime}}\Psi^{\prime}\varphi \, ; \, \phi)\right ]
 \label{Sclass}
 \end{eqnarray}
where $V_{cl}(\varphi \, ; \, \phi )$ is the classical TMT effective 
potential
for the matter (Higgs) scalar field $\varphi$ in the presence of the 
 background inflaton field $\phi$:
\begin{equation}
V_{cl}(\phi \, ; \, \varphi)=P(\varphi)\frac{M^{4}+V_{1}(\phi)}{2V_{2}(\phi)}
e^{\alpha\phi/M_{p}};
\label{Vcl}
\end{equation}
$L_{kin}(\overline{\Psi^{\prime}}, \Psi^{\prime}, A_{\mu})$ is the 
standard kinetic term for the fermion field in a Riemannian space-time
with metric $g^{\prime}_{\mu\nu}$, including also the gauge coupling to
the vector field $A_{\mu}$. And finally, the TMT effective "Yukawa 
coupling type" term 
$L_{Yuk}(\overline{\Psi^{\prime}}\Psi^{\prime}\varphi \, ; \, \phi)$
is
\begin{equation}
L_{Yuk}(\overline{\Psi^{\prime}}\Psi^{\prime}\varphi \,;\, \phi)=
-\frac{h}{\sqrt{2}}\overline{\Psi}^{\prime}\Psi^{\prime} 
\varphi \frac{e^{\gamma\phi/M_{p}}}{\chi^{3/2}}= 
-\frac{h}{4}\overline{\Psi^{\prime}}\Psi^{\prime}\varphi
\left[\frac{M^{4}+V_{1}(\phi)}{V_{2}(\phi)}\right]^{3/2}
e^{\gamma\phi/M_{p}}.
 \label{LYuk}
\end{equation}

\subsection{Massless scalar electrodynamics model in the
cosmological background and SSB}

Up to now the function $P(\varphi)$ was unspecified. Ignoring here 
technical questions (in particular, the question of
renormalizability that requires a non-minimal coupling  
$\eta R|\varphi |^{2}$ ), let us attract attention
to  a quantum effective  potential when choosing 
$P(\varphi)=\frac{\overline{\lambda}_{0}}{4!}\varphi ^{4}$, \quad 
$\overline{\lambda}_{0} =const$.
This means that (ignoring the fermion field), we are 
dealing  with  massless scalar
electrodynamics in curved space-time where the classical potential 
(the tree approximation)
is given by
\begin{equation}
V_{cl}(\varphi \, ; \, \phi)=\frac{\lambda_{0}(\phi)}{4!}\varphi ^4
\label{tree}
\end{equation}
and $\lambda_{0}(\phi)$ depends on the background field  $\phi$:
\begin{equation}
\lambda_{0} (\phi) =\overline{\lambda}_{0}k(\phi);\qquad 
k(\phi)\equiv\frac{M^{4}+V_{1}(\phi)}{2V_{2}(\phi)}
\ e^{\alpha\phi /M_{p}}.
\label{lambda0}
\end{equation}

Numerical estimations of $k(\phi)$ in the models of Sec. IV.B give the
following results:  $0<k(\phi)<3.5$ for model 1; \, $0<k(\phi)<
1.2\cdot 10^{-8}$ for model 2;\, $0<k(\phi)<3\cdot 10^{-7}$ for model 3.
In all models $k(\phi)$ asymptotically approaches zero as $\phi\rightarrow
\pm\infty$. Thus, in all cases 
$\lambda_{0} (\phi)$ is of the same order or less than 
$\overline{\lambda}_{0}$. 

The computation technics of the effective potential for the "massless"
scalar electrodynamics  in the one-loop approximation
is well-known issue\cite{CW}. However, the problem we
study here is not quite usual: the quartic coupling "constant"
depends actually on the cosmic time via the inflaton field $\phi$. Taking
 into account that in a course of its  evolution, the classical
field
 $\phi$ remains practically constant during a typical time of quantum 
matter fields fluctuations,  it is natural to consider the problem 
in the adiabatic approximation. Therefore computing
the effective potential we can regard $\lambda_{0} (\phi)$ as a constant.
Then the computation becomes quite standard. The only additional issue we 
have to clear up is a possible physical effect that the adiabatically 
changing $\lambda_{0} (\phi)$ might be on the $\varphi$-efffective 
potential .

One can check that the first point where we encounter necessity to decide
this problem, is the renormalization procedure. In fact, performing
calculations with the bare coupling constant $\lambda_{0}$ we have no 
need to think about its adiabatic $\phi$ dependence. But when we turn
to the use of the renormalized (finite) parameter $\lambda$ defined by
$ \lambda_{0}=\lambda +\delta\lambda$ where  
$\delta\lambda$ is the counter term (which, as one knows, is divergent
in perturbation theory), we have to take into account a possible
$\phi$-dependence of the effective $\lambda$ . 

The  vector boson loops contribution to the effective potential 
in the one-loop approximation has
the order of $e^4$ and {\em does not depend on} $\phi $ 
 (see Eq. (\ref{Sclass})). Therefore, just as in the standard scalar
electrodynamics, one can assert that in spite of possibility for 
$\lambda_{0}(\phi)$ to be very small, the effective $\lambda (\phi)$ can
not be too small. On the other hand it is important also that 
$\lambda (\phi)$ can not be large: since
$\phi$-dependence of $\lambda_{0}$ acts in the direction of decrease
in comparison with $\overline{\lambda}_{0}$, there are no reasons for
a possible $\phi$-dependence of $\lambda $ to act in the opposite 
direction.  

The scalar loops contribution has the order of $\lambda^{2}$. Therefore,  
in the same way as in the 
standard scalar electrodynamics, in the one-loop approximation, one 
can neglect the scalar loops contribution with
respect to the vector boson loops contribution. 
 
The  one-loop effective potential for the scalar field $\varphi$ 
evaluated at the fixed value of the
background inflaton field $\phi =\phi_{1}$ can be written in the form
\begin{equation}
V_{eff}(\varphi ;\phi_{1})=\frac{\lambda(\phi_{1})}{4!}\varphi^{4}+
\frac{3e^{4}}{(8\pi)^{2}}\varphi^{4}\left(\ln\frac{\varphi^{2}}
{\mu^{2}}-\frac{25}{6}\right),
\label{effpotvarphi}
\end{equation}
where 
\begin{equation}
\lambda(\phi_{1})=\frac{d^{4}V_{eff}}{d\varphi^{4}}\vert_{\varphi
=\mu}.
\label{lambdadefin}
\end{equation}

Let us assume that $\phi_{1}$ is the value of the
background inflaton field where $\lambda(\phi)$ has a maximal possible
magnitude (but it is still small!).
Suppose also that the renormalization mass $\mu$ is chosen such that 
$\lambda(\phi_{1})\sim e^{4}$. This can be always done
as  is well known from the renormalization group analysis\cite{CW}.
The final form of the effective 
potential  
\begin{equation}
V_{eff}(\varphi ;\phi_{1})=
\frac{3e^{4}}{(8\pi)^{2}}\varphi^{4}\left(\ln\frac{\varphi^{2}}
{\upsilon^{2}}-\frac{1}{2}\right),
\label{effpotvarphi1}
\end{equation}
 is determined in terms of two {\em free} parameters: renormalized
gauge coupling constant $e$ and VEV $<\varphi>=\upsilon$.

To verify whether the change of the value of the background inflaton 
field $\phi$ has some physical consequences, let us suppose that we 
want to repeat the same computation of the one-loop effective potential 
at an another fixed value of the
background inflaton field $\phi =\phi_{2}$ where the order of 
magnitude of $\lambda(\phi_{2})$ is less than $e^{4}$, if we take the same
renormalization mass $\mu$.
 According to  results of
the renormalization group analysis\cite{CW}, one can move 
$\lambda(\phi_{2})$ to the magnitude of the order of $e^{4}$ by
a change in the renormalization mass  that {\em does not change the order 
of 
magnitude of} $e$. This can be always done if $\lambda(\phi_{2})$ is small. 
Then the computation of 
the one-loop effective potential at $\phi =\phi_{2}$ {\em in the same 
approximation} leads to 
the same effective potential as it was at $\phi =\phi_{1}$ , 
Eq. (\ref{effpotvarphi1}), {\em with the same 
order of magnitude of the free parameter} $e$. One can conclude therefore
that in the used approximation, the $\phi$-dependence of $\lambda$ has no
physical effect.

I will ignore on this stage of investigation the fermion loops 
contribution into the $\phi$  effective 
potential . The nonminimal coupling of the Higgs field 
$\varphi$ to curvature, which
appears in the quantum effective action in curved space-time\cite{BO},   
might have some interesting but, most likely, weak enough effect, and 
this question exceeds the limits of the present paper. 

So, for the usual enough form of the function $P(\varphi)$,
 we obtain, in a cosmological background,  the effective 
quantum potential for the scalar (Higgs) field $\varphi$ typical for 
 gauge  theories with dynamical symmetry breaking.
Notice again that the term 
$\frac{e^{2}}{2}\varphi^{2}A_{\mu}A_{\nu}g^{\prime\mu\nu}$ in 
Eq.(\ref{Sclass})
 does not depend on the inflaton 
field $\phi$. 
Thus, SSB and Higgs phenomenon occur in a standard way.

\subsection{Yukawa coupling type term and fermion mass}

As a result of SSB, the Yukawa coupling type term (\ref{LYuk})
(see also Eq. (\ref{PsiEin})) 
produces the TMT effective fermion mass $m_{f} $ depending on the 
inflaton field $\phi$:
\begin{equation}
m_{f}  =m_{f} (\phi)=
\frac{h}{4} \upsilon
\left[\frac{M^{4}+V_{1}(\phi)}{V_{2}(\phi)}\right]^{3/2}
e^{\gamma\phi/M_{p}}.
\label{mupsi}
\end{equation}

For   $\phi > M_{p}$ (the region corresponding to the late 
universe ),
 the fermion mass becomes
\begin{equation}
m_{f}^{(late)} \simeq m_{f}^{(0)}\, e^{-[3(\beta -\frac{1}{2}\alpha) -
\gamma ]\phi/M_{p}}, \qquad
m_{f}^{(0)}=2h\upsilon \left(\frac{m_{1}}{m_{2}}\right)^{6}, \qquad as
\qquad \phi > M_{p}
\label{muphilarge}
\end{equation}
We see that in the late universe  the fermion mass  approaches the nonzero
 constant $m_{f}^{(0)}$ if  
 \begin{equation}
\gamma = 3(\beta -\frac{\alpha}{2}).
\label{gamma}
\end{equation}

Notice that if $\gamma$ indeed satisfies the relation (\ref{gamma})
then with the choice as in Sec. IV (i.e. $\alpha =6$ and $\beta =7$), 
we obtain
$\gamma =12$ that is close to the value of $\frac{3}{2}\beta =10.5$ 
dictated by the symmetry (\ref{fullglobal}) (see discussion after Eq.
(\ref{SYuk, original})) \footnote{The value $\gamma =12$ 
is as close to $\frac{3}{2}\beta =10.5$ as $\beta =7$
is close to $\alpha =6$. }. So in the framework of our
working hypothesis about approximate symmetry (\ref{fullglobal}) one can 
ensure a {\em successful mass generation for fermions in the present
cosmological epoch in a way typical for the standard model} and, at the 
same time, one to keep the direct coupling , Eq.(\ref{LYuk}), of fermionic
 matter to inflaton field (compare this with Wetterich's model 
\cite{Wett}). It is very interesting that only if the relation (\ref{gamma})
holds, the equations of motion in the Einstein frame (see subsection VID)
display the asymptotic restorstion of the symmetry $\phi\rightarrow
\phi +const$ as $\phi\rightarrow\infty$.

One has to notice that a formal generalization of the toy (abelian) model
 we 
study here,  to a non-abelian one (like $SU(2)\times U(1)$ or $SU(5)$)
can be performed straightforward. Then we have to worry about scales
of the particles mass generated as a result of SSB. In this connection it would be
interesting to estimate the order of magnitude of the fermion mass 
in the present universe that one could expect on the basis of 
Eqs. (\ref{muphilarge}) and (\ref{gamma}).
With the mass parameters $m_{1}$ and  $m_{2}$  of the models of Sec. IV.B 
(that implies $m_{1}/{m_{2}}=10^{-2}$ for model 1 and
 $m_{1}/{m_{2}}=10^{-3}$ for models 2 and 3 ) and with 
$\upsilon\sim 10^{2}\,GeV$ , estimations give too small values for fermion
mass at the late universe: $m_{f}^{(0)}\sim h\cdot 10^{-1}eV$ in model 1 
and
$m_{f}^{(0)}\sim h\cdot 10^{-7}eV$ in models 2 and 3. This is because of 
the 
presence in (\ref{muphilarge})  of the very small factor 
$(\frac{m_{1}}{m_{2}})^{6}$. 

Masses of the vector bosons,
as it was explained at the end of the previous subsection, do not depend
on the inflaton field $\phi$ and their values are defined  as in 
the standard gauge unified models. For the mass generation of fermions 
we have more 
freedom than in the standard models. 
According to
the basic ideas of the model developed in the present paper, the general
structure of Eq. (\ref{mupsi}) for masses of fermions is the same for 
field theory models
with different symmetry groups. The only free parameters, besides the 
inflaton field $\phi$, are the VEV $\upsilon$ of the appropriate scalar
boson and $\gamma$. If the values of $\gamma 's$ are determined by Eq. 
(\ref{gamma}) then masses of all fermions in the present universe are 
constants. If however, the parameter $\gamma$ corresponding to some of 
the fermions is such that
$3(\beta -\frac{1}{2}\alpha) -\gamma $ is very small but non-zero,
then 
$m_{f}$ becomes slow $\phi$-dependent according to Eq. (\ref{muphilarge})
even at the late universe. Namely, since in the quintessence model
 with exponential potential (\ref{quint}),
the inflaton field $\phi$ changes\cite{Wett} in cosmic time as 
$\phi\propto\frac{M_{p}}{\beta -\alpha}\ln t$, \quad we obtain that
 $m_{f}(\phi (t))$ will change in such a case as
$t^{-[3(\beta - \frac{1}{2}\alpha) -\gamma]/(\beta -\alpha )}$.
If $|3(\beta - \frac{1}{2}\alpha) -\gamma|\ll \beta -\alpha $
(in models of Sec.IV.B this means $|12-\gamma |\ll 1$), the rate of change
of $m_{f}$ might be very small in the present universe. Depending on the
sign of $3(\beta - \frac{1}{2}\alpha) -\gamma$, that should not be the 
same for all fermions, $m_{f}$ could be either increasing or decreasing.  
Notice that the case $3(\beta - \frac{1}{2}\alpha) -\gamma\leq 0$
 corresponds in some sense to the model studied by Wetterich\cite{Wett}.

Concerning the very early universe, that is for $\phi < -M_{p}$, 
one can see that the model predicts for the  TMT effective
 fermion mass, Eq. (\ref{mupsi}), to be extremely
small: $m_{f}\rightarrow 0$ as $\phi\rightarrow -\infty$. For example, in 
 model 3 of Sec. IV.B, 
$m_{f}\simeq h\upsilon 10^{-2}e^{-\gamma |\phi |/M_{p}} \quad as
\quad \phi < -M_{p}$. 
At the same time, the gauge 
coupling of 
$\Psi^{\prime}$ to $A_{\mu}$ (see Eq. (\ref{PsiEin})) is the standard one 
and,
in particular, it does not depend on the inflaton field $\phi$.  
\footnote{This can be an interesting example of the model\cite{TDLee}
 of massless spinor 
electrodynamics realized as the limit of a massive theory
as $\phi\rightarrow -\infty$.}

\subsection{The long-range force problem }

The r.h.s. of Eq. (\ref{phiEin}) describes a model with the direct coupling
of the inflaton to fermionic matter. For all models of Sec. IV.B, at the 
present universe, i.e.
 in the {\em quintessential region} 
(see Sec. IVC, item 3), the effective Lagrangian of this coupling takes the form 
\begin{equation}
L_{eff, present}^{(Yuk)}=
-\gamma\frac{m_{f}^{(late)}}{M_{p}}
\overline{\Psi^{\prime}}\Psi^{\prime}\phi .
\label{effYuk}
\end{equation}

Assuming the condition (\ref{gamma}) for constancy of $m_{f}^{(late)}$ 
 and with the choice $\beta =7$, 
$\alpha =6$, we get that the coupling constant of the present day 
effective Yukawa coupling of the inflaton to fermion is 
$12\frac{m_{f}^{(0)}}{M_{p}}$. Existing of such coupling 
would produce a too strong  scalar long-range force. Fortunately, TMT 
gives us additional tools that allow to solve this problem.

In the model\cite{G1} with the only {\em spontaneous breaking} of the global 
continuous symmetry (\ref{fullglobal}),  Guendelman studied the case where
the direct fermion-inflaton couplings similar to (\ref{SYuk, original})
 present in the original TMT action both with the measure $\Phi$ and 
with the measure $\sqrt{-g}$. In such a model the constant fermion mass
is also achieved\cite{G1}. Having this idea in mind, let us modify our model
 (\ref{L1gr+mat}),
(\ref{L2gr+mat}) with the explicit breaking of the symmetry 
(\ref{fullglobal}), by including an additional Yukawa coupling type term 
which enters into the action 
with the measure $\Phi$
\begin{equation}
 \tilde{S}_{Yuk}=
-\tilde{h}\int\overline{\Psi}\Psi |\xi |e^{\tilde{\gamma}\phi/M_{p}}
\Phi d^{4}x.
\label{SYuk tilde original}
\end{equation}
For this term to be invariant under transformations 
 (\ref{fullglobal}), the parameter 
 $\tilde{\gamma}$ must be $\tilde{\gamma}=\frac{1}{2}\beta <\alpha$.
 The magnitude
of $\tilde{\gamma}$ preferable from the dynamical point of view will
 be discussed below.

One can check that in this modified model, the fermion mass at the 
late universe becomes
\begin{equation}
m^{(late)}_{f, modified} \simeq \upsilon  
\left(\frac{m_{1}}{m_{2}}\right)^{2}
\left[2h\left(\frac{m_{1}}{m_{2}}\right)^{4}
               e^{-[3(\beta -\frac{1}{2}\alpha) -\gamma ]\phi/M_{p}}+
   \tilde{h}e^{-(\beta -\frac{1}{2}\alpha -\tilde{\gamma})\phi/M_{p}}
\right]   
 \qquad as
\qquad \phi > M_{p}.
\label{muphilargemodified}
\end{equation}
The constancy of $m^{(late)}_{f, modified}$ is achieved now if the condition
\begin{equation}
 \tilde{\gamma}=\beta -\frac{1}{2}\alpha
\label{gammatilde}
\end{equation}
holds together with (\ref{gamma}). For $\beta =7$ and $\alpha =6$, 
the constancy of the fermion mass at the late universe implies that
$\tilde{\gamma}=4$ which is as close to $\tilde{\gamma}=\frac{1}{2}\beta =
3.5$ as $\alpha$ is close to $\beta$.

With the conditions for constancy of the fermion mass at the late universe,
Eqs. (\ref{gamma}) and (\ref{gammatilde}) , 
the modified effective Yukawa coupling of 
the inflaton to fermionic matter takes now the form 
\begin{equation}
L_{eff, present}^{(Yuk, modified)}=
-\frac{\upsilon}{M_{p}}
\left(\frac{m_{1}}{m_{2}}\right)^{2}(\beta -\frac{1}{2}\alpha)
\left[6h\left(\frac{m_{1}}{m_{2}}\right)^{4}+\tilde{h}\right]
\overline{\Psi^{\prime}}\Psi^{\prime}\phi.
\label{effyukmod}
\end{equation}

We see that in the modified model there exists a possibility to prevent 
the appearance of such
danger interaction. To realize this opportunity we have to require 
\begin{equation}
\frac{\tilde{h}}{h}=-6\left(\frac{m_{1}}{m_{2}}\right)^{4}.
\label{tildehh}
\end{equation}
This is actually strong enough tuning since for instance, in the context 
of models 2 and 3 of Sec. IV.B, it implies $|\tilde{h}/h| \sim 10^{-12}$.
If we recall that $\tilde{h}$ and $h$ are the Yukawa type coupling 
constants of the Higgs scalar to fermion, it appears to be surprisingly
that their ratio has to be of the order 
of magnitude that shows the degree of the 
hierarchy problem in GUT: $m_{W}/m_{X}\sim 10^{-12}$.

With conditions (\ref{gamma}),(\ref{gammatilde}) and (\ref{tildehh}),
 the fermion mass at the late universe becomes
\begin{equation}
m^{(late)}_{f, modified} = 
\frac{2}{3}\tilde{h}\upsilon \left(\frac{m_{1}}{m_{2}}\right)^{2}
 \qquad as
\qquad \phi > M_{p}
\label{muphilargemodfinal}
\end{equation}

A possible relation of the discussed question to the hierarchy problem
in GUT, as well as other problems that appear in the attempts 
 to generate a realistic unified gauge theories in
the context of TMT, will be studied elsewhere.

\section{Discussion and Conclusion}

Before summarizing and discussing main results of this paper I would 
like to stress again that {\em the first impression that the studied models 
belong to a sort of a scalar-tensor theories, is wrong}. 
The
ratio of two measures, that is the scalar field $\chi$, Eq. (\ref{chi}),
is the only object entering into equations of motion and carrying
information about the measure $\Phi$ degrees of freedom. If we restrict 
ourselves by
models where $L_{1}$ is linear in the scalar curvature (see Eqs. 
(\ref{S}), (\ref{totac}) and (\ref{L1gr+mat})) and $L_{2}$ does not contain 
curvature,
then in the first order formalism, the constraint appears that determines
$\chi$ in terms of matter fields (see Eqs. (\ref{c}) or (\ref{con+matt})).
  This means that in such models the
scalar field $\chi$ does not carry an independent degree of freedom.  
All deviations from the Einstein or Einstein-Cartan theory existing in the
original variables are caused by derivatives of $\sigma\equiv\ln\chi$ and
they disappear in the new variables obtained by the conformal 
transformations (\ref{ct}) or (\ref{ctferm}). By an appropriate choice of
$L_{1}$ and $L_{2}$ one can provide that {\em all equations of motion 
in the new variables have canonical GR form of equations for
 gravity and matter fields}.
All novelty is revealed only in an unusual structure of the effective
potentials and interactions. And just this novelty enables to solve a
number of problems (questions 1-5 of Introduction) most of 
 which in the framework of GR require fine tuning. 
 
\medskip
\paragraph{Towards a resolution of the cosmological constant problem.}

Let us return for the moment to the simple model of Sec. II. If one takes
\cite{GK4} $V_{2}(\phi)\equiv -\Lambda =const$ that in GR would correspond
\footnote{Taking into account our definition of $V_{2}(\phi)$, 
Eq. (\ref{totac}), one should notice that the positive $V_{2}^{0}$ 
corresponds
to a negative cosmological constant $\Lambda =-V_{2}^{0}$ in GR if the term 
$\int V_{2}(\phi)\sqrt{-g}d^{4}x$ would appear in the GR action. For
constructing  models 1 - 3 of Sec. IV.B, the positive
 definiteness of $V_{2}(\phi)$ (and therefore the condition 
$V_{2}^{0}> 0$) 
was one of the basic assumptions.}
 to a model with a cosmological
constant $\Lambda$, then we see that
 the greater $|\Lambda | $ we admit,
 the smaller TMT effective potential, Eq.(\ref{VW}), we obtain in the 
Einstein picture. This is a direct result of existence of two measures and 
two Lagrangians in the original TMT action, Eq. (\ref{totac}).
We see that TMT turns over  our intuitive ideas based on
our experience in field theory. 

The resolution of the cosmological constant problem in models studied in
Refs.
\cite{GK3}-\cite{G1} was based on the assumption 
that a
cosmological scenario belongs to the first class (see Sec. II). In the
context of such types of scenarios, those TMT models predict that if 
$V_{2}(\phi)$ is positive definite, the stable vacuum with zero energy 
density
is realized without any sort of fine tuning at a finite value of $\phi =
\phi_{0}$ where $V_{1}(\phi_{0})+sM^{4}=0$. 

In this paper, we studied 
cosmological scenarios of the second class (see Sec. II) where the 
true vacuum state  is
realized asymptotically as $\phi\rightarrow\infty$. This naturally leads us
to a need to apply to a quintessence model of the late universe. However,
in contrast to quintessence models studied in the framework of GR or 
Brans-Dicke type models, in TMT we have a {\em new option}: one can choose
the
prepotentials $V_{1}$ and $V_{2}$ {\em increasing at the late universe} 
(that is 
as $\phi >M_{p}$). If $V_{1}^{2}/V_{2}$ approaches zero as 
$\phi\rightarrow\infty$,
then the TMT effective potential (\ref{VW}) asymptotically approaches zero
at the late universe.  One can adjust degrees of growth of $V_{1}$ and 
$V_{2}$ in such a way that the TMT effective potential $U(\phi)$ will have
 a desirable flat  shape as $\phi\rightarrow\infty$. Unbounded growth of
$V_{2}$ as $\phi\rightarrow\infty$ allows adding to $V_{2}$ any 
 constant   $V_{2}^{0}$  without altering  $U(\phi)$ for 
$\phi$ large 
enough (remind that appearance of an additive constant in $V_{1}$ does 
not affect equations of motion at all). This is actually what we have seen
in Sec. III. 
If  appearance of the appropriate
term $\int V_{2}^{0}\sqrt{-g}d^{4}x$ in the action 
is a result of quantum vacuum fluctuations then we can conclude that in the
framework of the described approach to constructing a quintessence model of
the late universe, {\em TMT solves the cosmological constant problem}.

However, the  impression  that the described technical details of the 
approach to the resolution of 
the cosmological 
constant problem in TMT settles a question,  is premature.
 One should remind that the last statement about resolution of 
the cosmological constant problem implies validity of  one
more basic
conjecture formulated in Introduction (after Eq. (\ref{Fi})) and used in
all models of the present paper:  Lagrangians  $L_{1}$ and $L_{2}$ 
in the original action (\ref{S}) do not depend on the
measure $\Phi$ degrees of freedom. In cases when this conjecture is invalid,
the cosmological constant problem in TMT can turn into a very nontrivial
issue.  In fact, till  the fundamental theory remains unknown, one can
not
be sure that the postulated general structure of TMT survives after
 quantum corrections are taken into account. If it will 
turn out that the quantum effective action corresponding to the original
theory
(\ref{S}), contains the term
$-\int\Phi\chi\Lambda_{eff}d^{4}x$, then in the Einstein frame the latter 
will generate the real cosmological constant $\Lambda_{eff}$.
This possibility was studied in Ref. \cite{GK4} (see Sec. VI therein) where 
a way to prevent the appearance of such a danger term was also discussed.
The idea, briefly, is the following.
If instead of the antisymmetric tensor field $A_{\mu\nu\lambda}$, the 
measure $\Phi$ is defined by means of four scalar measure fields 
$\varphi_{a}, (a=1,2,3,4)$, 
\begin{equation}
\Phi \equiv \varepsilon_{a_{1}a_{2}a_{3}a_{4}}
\varepsilon^{\mu\nu\lambda\sigma}
(\partial_{\mu}\varphi_{a_{1}})
(\partial_{\nu}\varphi_{a_{2}})
(\partial_{\lambda}\varphi_{a_{3}})
(\partial_{\sigma}\varphi_{a_{4}}),
\label{Fia4}
\end{equation}
then the action (\ref{S}) with  $\varphi_{a}$ - independent
$L_{1}$ and $L_{2}$, is invariant, up to an integral of a total divergence,
under transformations
$\varphi_{a}\rightarrow\varphi_{a}+f_{a}(L_{1})$ where $f_{a}(L_{1})$ are 
arbitrary differentiable functions of $L_{1}$. An appearance of the danger 
term
$-\int\Phi\chi\Lambda_{eff}d^{4}x$ in the action would break this local
symmetry. Thus, this additional, local symmetry can prevent a generation
of the real cosmological constant  by quantum corrections to TMT
if no anomaly appears.

\medskip
\paragraph{Resolution of the flatness problem of 
the quintessential  potential.}

The mechanism for the resolution of the flatness problem of 
the quintessence  potential in TMT (question number 2 of Introduction) is
actually the same as the one used for the resolution of 
the cosmological constant problem. Since the TMT effective   potential 
$U(\phi)$
takes a quintessence form as $\phi\rightarrow\infty$ due to
 the unbounded growth of the leading terms of
the pre-potentials $V_{1}$ and $V_{2}$, appearance of any subleading terms 
(including terms generated by quantum corrections) in
$V_{1}$ and $V_{2}$ can not alter the shape of $U(\phi)$ as $\phi$ is 
large enough. There is no any need for coupling constants and mass parameters
 of the subleading terms to be very small. This is in fact the TMT answer 
  the question raised by Kolda and Lyth\cite{KLyth}. 
 
\medskip
\paragraph{"Quintessential inflation" type potential (satisfying also
the cosmological nucleosynthesis constraint)
 obtained without fine tuning.}   

Two basic ideas has been used in this paper  to demonstrate that
 TMT enables to answer questions 3 and 4 of Introduction .
The fundamental role belongs to the first
idea that in the limit $\phi\rightarrow\infty$, the effective theory has to 
become invariant under  shifts $\phi\rightarrow\phi +const$. A basis for 
this idea is the observation that if we want that  effective theory would
describe a quintessence as $\phi\rightarrow\infty$, the effective potential
has to become flat as $\phi\rightarrow\infty$. 

As it was shown by 
Guendelman\cite{G}, \cite{G1} the role of the global continuous symmetries 
$\phi\rightarrow\phi +const$ in
TMT belongs to transformations (\ref{st1}), (\ref{st3}) in the absence
of fermions or (\ref{fullglobal}) in the presence of fermions: in terms of
the dynamical variables used in the Einstein frame, these transformations
are reduced just to shifts of $\phi$ parametrized as in Eq. (\ref{st3}).
In models of Ref. \cite{G}, where the exponential form for the pre-potentials
(\ref{pot12}) with $\alpha =\beta$ has been used, the global  symmetry 
(\ref{st1})-(\ref{st3}) is {\em spontaneously broken}.
 And although this symmetry is restored as $\phi\rightarrow\infty$, it
is impossible in the framework of such a model to realize a quintessence 
scenario at $\phi >M_{p}$.  

We have seen in the present paper that if a small {\em explicit violation}
 of the
global continuous symmetries (\ref{st1}), (\ref{st3}) is present in the
TMT original action (\ref{totac}) with the exponential form of the 
pre-potentials (\ref{pot12}), then the TMT effective potential $U(\phi)$,
Eq. (\ref{U1}), can be a suitable candidate for a quintessence model 
as  
$\beta\phi\gg M_{p}$. The smallness of the explicit symmetry breaking is
formulated as a smallness of the dimensionless parameter
 $(\beta -\alpha )/\beta$, see Eq. (\ref{fla<b}). 

In the absence of a knowledge about the structure of the fundamental theory
 and without
any information  about a mechanism leading to an explicit violation of the
global continuous symmetry (\ref{st1}), (\ref{st3}),  the quantity
$(\beta -\alpha )/\beta$ is the only small parameter that can be used in 
attempts  to modify the action with simple exponential form of the 
pre-potentials (\ref{pot12}), with the aim to give rise to quintessential 
inflation type models. This can be done by adding terms that disappear as
$(\beta -\alpha )/\beta$ tends to zero. This means that coupling constants
in such additional terms have to be proportional to some positive power of
this small parameter.

 The second basic idea is that
in the limit $(\beta -\alpha )/\beta\rightarrow 0$ (which leads us to the
fundamental theory) the only mass parameter of the theory is the Planck
mass $M_{p}$. This means that  the dimensional coupling constants of
 the  symmetry breaking terms have to be  powers of the mass parameters
$m$ of the form $m=[(\beta -\alpha )/\beta]^{n}M_{p}, \quad n>0$. 

In the probe models studied in Sec. IV, we have chosen 
just for illustration $\beta =7$, 
$\alpha =6$ and hence $(\beta -\alpha )/\beta =1/7$. Proceeding in the
described above way,
we reveal a remarkable feature of TMT: it is possible to achieve quite
satisfactory quintessential inflation type models 
(see models 1-3 of Sec. IV) where for  adjustment of the parameters
  it is enough to 
use only mass parameters  of few orders  less than $M_{p}$. We 
interpret this fact as the absence of a need of a fine tuning.

Besides of the generation of the well-defined inflationary and
quintessential regions of the TMT effective potential $U(\phi)$,
one more remarkable result consists in the fact that the post-inflationary
region of $U(\phi)$ has the exponential form $\propto exp(-a\phi/M_{p})$
 with 
variable $a$, Eq. (\ref{vara}). This allows to single out a region of 
$U(\phi)$, where a familiar approach\cite{Wett} to a 
resolution of the problem with the  cosmological
 nucleosynthesis constraint is realized without any additional assumption.

\medskip
\paragraph{Resolution of the problems related to a possible direct coupling
of the inflaton field to usual  matter.}

As to the question number 5 of Introduction,  the answer is quite clear:
if the terms of the form $f_{i}\frac{\phi}{m}{\cal L}_{i}$, describing
direct couplings 
of the inflaton field to the usual matter (see Ref. \cite{Carroll}), break 
the
global continuous symmetry (\ref{fullglobal})  they
could appear in the original TMT action  with small coefficients
$f_{i}\propto [(\beta -\alpha )/\beta ]^{n}, \quad n>0$.

A direct coupling of the inflaton  to
 {\em a fermionic matter} is of a special interest. In the modified 
model  studied in Sec. VI.H,
such a coupling enters to the original action in the form of two Yukawa 
coupling type terms, 
Eq. (\ref{SYuk, original}) and (\ref{SYuk tilde original}).
 The unbounded encrease of $V_{1}$ and $V_{2}$ at the late universe works
again in the desirable direction: the contributions of 
the Yukawa coupling type terms to the constraint (\ref{con+matt}) are 
negligible compared to $V_{1}$ and $V_{2}$. As  we have seen, by adjustment
of the parameters of
the Yukawa coupling type interactions one can provide  the
presence of the direct coupling of fermionic matter to inflaton without 
observable  effects at the late universe:  
 the fermion mass approaches constant and the correspondent
long-range force disappears as $\phi\rightarrow\infty$.

It is worthwhile to notice here that the form of the Yukawa coupling type
interactions
(\ref{SYuk, original}) and (\ref{SYuk tilde original}) might be
generalized without altering the
results obtained for the late universe. In fact, if for example one
modifies
the  interactions of the form $\propto\overline{\Psi}\Psi 
e^{\gamma\phi/M_{p}}\sqrt{-g}$ and
$\overline{\Psi}\Psi 
e^{\tilde{\gamma}\phi/M_{p}}\Phi$ considered in Sec. VI.H, 
by adding the direct couplings of the form
$\propto\overline{\Psi}\Psi\phi\sqrt{-g} $ and
$\overline{\Psi}\Psi\phi\Phi$ respectively
(which does not respect the global
continuous symmetry (\ref{fullglobal})  )
 this has no effect on the late universe
since the relative contributions of the adding terms are exponentially
suppressed as $\phi >M_{p}$ . At the early universe,
for instance as $\phi <0$,  modifications like this  could lead to
observable effects. Such
possibilities are additional tools given by TMT  for adjustment  the
field theory parameters to cosmological  constraints of the early universe.
One should stress that this is a merit of TMT  that {\em 
adjustment of  the
parameters determining  the early universe evolution can be performed
without
any direct influence on the field theory parameters important for the
 late universe}.

\medskip
\paragraph{Matter fields quantization.}

Quantization of usual matter fields in TMT has some specific problem.
In particular, fermionic field $\Psi$ in the model of Sec. VI
 contributes to the constraint, 
Eq. (\ref{con+matt}), and hence $1/\chi$ obtained by solving
(\ref{con+matt}), will  depend on $\overline{\Psi}\Psi$. In such a case,
equations of motion in the Einstein frame, (\ref{PsiEin}), (\ref{phiEin})
and (\ref{varphiEin})
would become very nonlinear. In Ref. \cite{GK4}, we have tried to avoid
this sort of problems  by starting from the original action that was
 very non-linear in  $\overline{\Psi}\Psi$. 

In the present paper, where the inclusion of the usual matter is studied
in the context of the models of Sec. IV.B, intended to describe 
quintessential-inflation scenario {\em without fine tuning}, the problem
of a non-linearity in matter fields does not appear. The reason is just due to
 a way that we  solve the cosmological constant and other fine tuning 
problems: the parameters of prepotentials
$V_{1}(\phi)$, $V_{2}(\phi)$ and the integration constant $M^{4}$ are chosen
such that the matter
fields contributions to the constraint (\ref{con+matt}) are negligible 
compared to $V_{1}(\phi)$, $V_{2}(\phi)$ and $M^{4}$. Then for 
$1/\chi$ we obtain the expression described by Eq. (\ref{1/chi}), the same 
as in the absence of the usual matter.
As a result of this, in the Einstein frame the usual matter fields equations 
in the background have canonical form and their quantization becomes a 
standard procedure.
 
\medskip
\paragraph{SSB without generation of the cosmological constant.}

Reverting to the cosmological constant problem, it is worthwhile to notice 
in the conclusion that if the  scalar (Higgs) field $\varphi$ obtains a 
non-zero VEV, Eq. (\ref{varphidefinition}), the appearance of a constant 
part 
in $P(\varphi)$ leads just to a redefinition 
of $m_{1}^{4}$ (see Eq. (\ref{con+matt}) ). It is very
 important that in models 1 - 3 of Sec. IV.B, \enspace $m_{1}^{4}$  has 
the order of
$(10^{-2}M_{p})^{4}$ or $(10^{-3}M_{p})^{4}$.
The correction we neglect
in the l.h.s. of (\ref{con+matt}) when replace it by (\ref{1/chi}),
becomes of the order of 
$Q(\tilde{\varphi})/m_{1}^{4}$ where $Q$ is a polynomial in 
$\tilde{\varphi}$ ($|\tilde{\varphi}|\ll \upsilon \lesssim m_{1}$ )
that satisfies the condition $Q(0)=0$. Thus if 
$|P(\upsilon)|<m_{1}^{4}$,
 then spontaneous braking of a gauge symmetry
 does not affect  the magnitude of the effective cosmological constant
(at the late universe)  imitated by the quintessential potential (\ref{quint}).

Another 
possibility appears if the whole term 
$m_{1}^{4}e^{\alpha\phi/M_{p}}$ in the pre-potential $V_{1}(\phi)$ 
is generated by SSB.     In such a case the quintessential potential becomes
\begin{equation}
U(\phi)\approx\frac{[P(\upsilon)]^{8}}{M_{p}^{4}}
e^{-2(\beta -\alpha)\phi/M_{p}}
\label{LambdaSSB}
\end{equation}

This is the TMT mechanism 
which together with the shape of $U(\phi)$ in the inflationary region
predicted by each of the models 1 - 3 of Sec. IV.B, provides a resolution 
of one of the most serious aspect of the
cosmological constant  problem\cite{Star}: the need 
of an enormous fine tuning of initial conditions in models with SSB in order
to satisfy the dual requirement of 'large $\Lambda $ in the past + small
$\Lambda$ at present'.

\section{Acknowledgments}    

I am grateful to S. de Alwis, R. Brustein,  A. Davidson and
 D. Owen for useful discussions at various stages of the work. I am 
especially indebted  to E. Guendelman for   attention and help
during the evolution of this paper.

\end{document}